\documentclass[a4paper,fleqn,usenatbib]{mnras}
\usepackage{ragged2e} 
\usepackage[T1]{fontenc}
\usepackage{ae,aecompl}
\usepackage{graphicx}	
\usepackage{amsmath}	
\usepackage{amssymb}	
\usepackage{cleveref}   
\usepackage{hyperref}   
\usepackage[countmax]{subfloat}
\usepackage{booktabs}
\newcommand{\RNum}[1]{\uppercase\expandafter{\romannumeral #1\relax}}

\title[]{Optical Variability of Three Extreme TeV Blazars}

\author[A. Pandey et al.]
{Ashwani Pandey$^{1}$\thanks{E-mail: ashwanitapan@gmail.com}, 
Alok C. Gupta$^{1}$\thanks{E-mail: acgupta30@gmail.com},
G. Damljanovic$^{2}$,
P. J. Wiita$^{3}$,
O. Vince$^{2}$,
\newauthor
and
M. D. Jovanovic$^{2}$
\\ 
\\
$^{1}$Aryabhatta Research Institute of Observational Sciences (ARIES), Manora Peak, Nainital -- 263001, India\\
$^{2}$Astronomical Observatory, Volgina 7, 11060 Belgrade, Serbia \\
$^{3}$Department of Physics, The College of New Jersey, 2000 Pennington Road, Ewing, NJ 08628-0718, USA\\
}

\date{Accepted XXX. Received YYY; in original form ZZZ}

\pubyear{2020}

\begin{document}
\label{firstpage}
\pagerange{\pageref{firstpage}--\pageref{lastpage}}
\maketitle

\begin{abstract}
We present the results of optical photometric observations of three extreme TeV blazars, 1ES 0229$+$200, 1ES 0414$+$009, and 1ES 2344$+$514, taken with two telescopes (1.3 m Devasthal Fast Optical Telescope, and 1.04 m Sampuranand Telescope) in India and two (1.4 m Milankovi\'{c} telescope and 60 cm Nedeljkovi\'{c} telescope) in Serbia during 2013--2019. We investigated their flux and spectral variability on diverse timescales. We examined a total of 36 intraday $R-$band light curves of these blazars for flux variations using the power-enhanced {\it F}-test and the nested ANOVA test. No significant intraday variation was detected on 35 nights, and during the one positive detection the amplitude of variability was only 2.26 per cent. On yearly timescales, all three blazars showed clear flux variations in all optical wavebands. The weighted mean optical spectral index ($\alpha_{BR}$), calculated using $B - R$ color indices, for 1ES 0229$+$200 was 2.09 $\pm$ 0.01. We also estimated the weighted mean optical spectral indices of 0.67 $\pm$ 0.01 and 1.37 $\pm$ 0.01 for 1ES 0414$+$009, and 1ES 2344$+$514, respectively, by fitting a single power-law ($F_{\nu} \propto \nu^{-\alpha}$) in their optical ({\it VRI}) spectral energy distributions. A bluer-when-brighter trend was only detected in the blazar 1ES 0414$+$009. We briefly discuss different possible physical mechanisms responsible for the observed flux and spectral changes in these blazars on diverse timescales.
\end{abstract}



\section{INTRODUCTION} \label{sec:intro}
Blazars constitute the most enigmatic class of radio-loud active galactic nuclei (RLAGNs) and are usually divided into BL Lacertae objects (BLLs) and flat-spectrum radio quasars (FSRQs). Blazars exhibit large amplitude flux and spectral variability across the whole electromagnetic (EM) spectrum, and in general have core-dominated radio structures. The EM radiation from blazars is predominantly non-thermal and shows strong linear polarization ($>$ 3 per cent) in radio to optical wavelengths. According to the unified emission model of RLAGNs, blazar's relativistic jets point in the direction of the observer line of sight within $\lesssim$10$^{\circ}$ \citep{1995PASP..107..803U} and the emitted radiation from the jet is affected by relativistic beaming, which implies a shortening of timescales by a factor $\delta^{-1}$, where $\delta$ is the Doppler factor. 

Blazars emit significant radiation in the complete EM spectrum which gives observers an opportunity to generate their spectral energy distribution (SED) from low-energy radio bands to extreme high energies upto $\gamma-$rays. SEDs of blazars are double-humped, and the first SED hump (i.e. low energy) peaks in Infra-red (IR) to X-rays while the second hump (i.e. high energy) peaks in $\gamma-$rays (from GeV up to TeV energies). The low energy part of SED is dominated by synchrotron emission from the relativistic jet while the high energy one is produced by inverse-Compton (IC) radiation within the frequently accepted leptonic scenario. However, the origin of the high energy hump is still under some debate, and hadronic as well as lepto-hadronic models have been proposed to explain it \citep[e.g.][]{2007Ap&SS.307...69B}.
Based on the SED first hump peak frequency, i.e., the synchrotron peak frequency $\nu_{\rm syn}$, blazars have often been classified into three subclasses: LSP (low synchrotron frequency peaked) $\nu_{\rm syn} \leq$ 10$^{14}$ Hz, ISP (intermediate synchrotron frequency peaked) 10$^{14} < \nu_{\rm syn} <$ 10$^{15}$ Hz, and HSP (high synchrotron frequency peaked) $\nu_{\rm syn} \geq$ 10$^{15}$ Hz \citep{2010ApJ...716...30A}. In a recent study with a very large sample of blazars, a slightly modified classification scheme of blazars was suggested by \cite{2016ApJS..226...20F}. According to this revised classification these subclasses of blazars are defined as: LSP still having $\nu_{\rm syn} \leq$ 10$^{14}$ Hz, but ISP being in the range 10$^{14} < \nu_{\rm syn} <$ 10$^{15.3}$ Hz, and so HSPs have $\nu_{\rm syn} \geq$ 10$^{15.3}$ Hz. Another subclass of blazar would be the extreme high frequency peaked BL Lac objects (EHBLs) in which the synchrotron peak frequency $\nu_{\rm syn}$ lies at  $> $ 1 keV  \citep[or $ > 10^{17}$ Hz;][]{2001A&A...371..512C,2019MNRAS.486.1741F}.  These EHBLs are a growing subclass of blazar whose synchrotron emission peaks at medium to hard X-ray energies and are therefore good candidates for TeV detection, as their upscattered IC photons would have exceptionally high energies.

TeV emitting blazars are so far rather rare and mostly belong to the HSP class. The first significant TeV emission from a blazar was detected from Mrk 421 at 0.5 TeV with the Whipple observatory $\gamma-$ray telescope \citep{1992Natur.358..477P}.  Until 2003, only six TeV blazars were known which were having confirmed detection of TeV emission from two different $\gamma-$ray telescopes \citep[e.g.,][for a summary of their properties]{2004NewAR..48..367K}. The ground and space-based $\gamma-$ray facilities developed and made operational in last about one-and-a-half decades e.g. {\it HESS} (High Energy Stereoscopic System), {\it MAGIC} (Major Atmospheric Gamma-ray Imaging Cherenkov), {\it VERITAS} (Very Energetic Radiation Imaging Telescope Array System), {\it Fermi}, etc., have made a revolution in TeV $\gamma-$ray astronomy and have discovered a significant  number of TeV emitting galactic and extragalactic sources which now include 73 blazars\footnote{http://tevcat.uchicago.edu/}. Out of these, 14 TeV sources have been cataloged as EHBLs \citep{2019MNRAS.486.1741F}. 

Blazars show detectable flux variations on diverse timescales across the EM spectrum. Blazar variability timescales range from a few minutes to years. AGN flux variation timescales from a few minutes to a day is variously known as micro-variability \citep{1989Natur.337..627M}, intra-day variability (IDV) \citep{1995ARA&A..33..163W} or intra-night variability \citep{1996MNRAS.281.1267S}. Flux variability on timescales from several days a to few months is commonly known as short timescale variability (STV) and variability timescales from months to several years can be called long timescale variability (LTV) \citep[e.g.][]{2004A&A...422..505G}.    

Blazar flux variability on these diverse timescales is an important tool to understand the emission mechanism. A puzzling issue is the blazar flux variability observed on IDV timescale. To try to elucidate the nature of the emission mechanism of blazars through variability on different timescales, we have run a project over the past 15 years and have reported our results in series of papers \citep[e.g.,][and references therein]{2008AJ....135.1384G,2008AJ....136.2359G,2012MNRAS.425.1357G,2016MNRAS.458.1127G,2017MNRAS.465.4423G,2010ApJ...718..279G,2012MNRAS.420.3147G,2012MNRAS.425.3002G,2015MNRAS.452.4263G,2015A&A...582A.103G,2015MNRAS.450..541A,2015MNRAS.451.3882A,2016MNRAS.455..680A,2015MNRAS.451.1356K,2017ApJ...841..123P,2018ApJ...859...49P,2019ApJ...871..192P,2020ApJ...890...72P,2018MNRAS.480.4873A,2019ApJ...884..125Z,2019ApJ...887..185S}. To continue these variability studies over diverse timescales in optical bands, we report here on three TeV $\gamma-$ray detected EHBLs namely, 1ES 0229$+$200, 1ES 0414$+$009 and 1ES 2344$+$514 \citep{2018MNRAS.477.4257C,2019MNRAS.486.1741F} which we observed from four ground-based optical telescopes (two in India and two in Serbia) during 2013 -- 2019.         

This paper is organized as follows: Section \ref{sec:data} gives an overview of the telescopes and photometric observations we used and the data reduction procedure. Analysis techniques we used to search for flux variability are discussed in Section \ref{sec:analysis}. Results of our variability study are given in Section \ref{sec:res}. A discussion and conclusions of our study are given in Section \ref{sec:diss}. 


\section{OBSERVATIONS AND DATA REDUCTION} \label{sec:data}

Optical photometric observations of three extreme TeV blazars, 1ES 0229$+$200, 1ES 0414$+$009, and 1ES 2344$+$514, were carried out using the standard Johnson-Cousin $BVRI$ filters from 2013 September 6 to 2019 November 4 with four ground-based telescopes, two in India and two in Serbia. The details of these four telescopes and the detectors used for our observations are given in Table \ref{tab:telescopes} in Appendix \ref{app_a}. The complete observation logs for these blazars are presented individually in Tables \ref{tab:obs02}$-$\ref{tab:obs23} in Appendix \ref{app_b}. The exposure times range from 280--300 s in $B$ band, 220--250 s in $V$ band, 150--200 s in $R$ band, and 80--120 s in $I$ band.

We performed optical photometric observations of these three TeV blazars during 2016--2018 using two Indian telescopes, the 1.3 m Devasthal Fast Optical Telescope (DFOT) and 1.04 m Sampuranand Telescope (ST). Both these telescopes have Ritchey$-$Chretien (RC) Cassegrain optics and are operated by Aryabhatta Research Institute of Observational Sciences (ARIES), Nainital, India. Observations with the 1.3 m DFOT were taken using Andor 2K CCD camera, while we observed these blazars with 1.04 m ST using a PyLoN CCD, except on 2016 November 8 and 9 when the 1.04 m ST was still equipped with a Tektronics 1K CCD. 

We also monitored these three blazars with 1.4 m Milankovi\'{c} telescope and 60 cm Nedeljkovi\'{c} telescope, located at Astronomical Station Vidojevica (ASV), Serbia. The 1.4 m telescope is equipped with am Andor iKon-L CCD, while an Apogee Alta  E47 CCD camera is mounted on the 60 cm telescope.  

The standard optical photometric data reduction procedure we employed on all observations taken with the 1.3 m DFOT and 1.04 m ST is discussed in detail in \cite{2019ApJ...871..192P, 2020ApJ...890...72P}. It involves cleaning the raw image frame through bias-subtraction, flat-fielding and cosmic-ray removal using standard routines of  IRAF\footnote{Image Reduction and Analysis Facility (IRAF) is distributed by the National Optical Astronomy Observatory, which is operated by the Association of Universities for Research in Astronomy (AURA) under a cooperative agreement with the National Science Foundation.}, followed by performing aperture photometry in DAOPHOT\footnote{Dominion Astronomical Observatory Photometry} \RNum{2} software to get the instrumental magnitudes of all the sources in the frame. The data obtained from Serbian telescopes were processed following the same reduction steps as in the case of Indian telescopes but using the MaxIM DL\footnote{\url{https://diffractionlimited.com/help/maximdl/MaxIm-DL.htm}} software package.

We used the finding charts provided by Landessternwarte Heidelberg-K{\"o}nigstuhl\footnote{\url{https://www.lsw.uni-heidelberg.de/projects/extragalactic/charts}} for our observations of extreme TeV blazars. During each observation, we observed two or more comparison stars present in the blazar field. We used the one star having brightness and color most comparable to that of the blazar to get the calibrated magnitudes of the blazar. 
Since all the observations of a particular source were calibrated using the same standard star in a particular filter, the cross-calibration for different instruments doesn't significantly affect the measured magnitudes on longer timescales. For intranight variability studies we performed observations using one single instrument on a particular night.

During our campaign, we performed multiple $R$ band observations of these extreme blazars for a total of 41 nights. However, to search for microvariations, we only selected nights with at least 10 data points. By applying this criterion, 14 nights of observations of 1ES 0229$+$200, 8 nights of  those of 1ES 0414$+$009, and 14 nights of those of 1ES 2344$+$514 are qualified for IDV analysis. 

\begin{table*}
\caption{Results of IDV analyses of extreme TeV blazars}            
\label{tab:var_res1}                   
\centering 
\resizebox{\textwidth} {!}{                     
\begin{tabular}{lccccccccccc}           
\hline\hline                		 
Blazar Name & Obs. date & Obs. start time & Band & \multicolumn{3}{c}{{\it Power-enhanced  F-test}} & \multicolumn{3}{c}{{\it Nested ANOVA}} &  Status \\
\cmidrule[0.03cm](r){5-7}\cmidrule[0.03cm](r){8-10}& dd-mm-yyyy & JD & & DoF($\nu_1$,$\nu_2$ ) & $F_{\rm enh}$ & $F_c$  & DoF($\nu_1$,$\nu_2$ ) & $F$ & $F_c$ &\\
\hline
1ES 0229$+$200 
& 24-10-2016  	&  2457686.31115	 & R  &  39,195 & 0.40 & 1.71	& 7, 32 & 1.59 & ~3.26 & NV \\ 
& 25-10-2016  	&  2457687.34194	 & R  &  31,155 & 2.04 & 1.81	& 5, 24 & 3.54 & ~3.90 & NV \\ 
& 26-10-2016  	&  2457688.32026	 & R  &  29,145 & 1.24 & 1.84	& 5, 24 & 4.21 & ~3.90 & NV \\ 
& 24-11-2016  	&  2457717.28042	 & R  &  40,200 & 0.30 & 1.69	& 7, 32 & 8.82 & ~3.26 & NV \\ 
& 25-11-2016  	&  2457718.23810	 & R  &  49,245 & 1.28 & 1.62	& 9, 40 & 4.90 & ~2.89 & NV \\ 
& 10-10-2018  	&  2458402.27049	 & R  &  41,205 & 0.89 & 1.68	& 7, 32 & 2.48 & ~3.26 & NV \\ 
& 28-12-2018  	&  2458481.05990	 & R  &  33,165 & 0.32 & 1.78	& 5, 24 & 1.71 & ~3.90 & NV \\ 
& 29-12-2018  	&  2458482.05678	 & R  &  46,230 & 1.26 & 1.64	& 8, 36 & 3.54 & ~3.05 & NV \\ 
& 28-08-2019  	&  2458724.42101	 & R  &  47,188 & 0.55 & 1.65 	& 8, 36 & 2.11 & ~3.05 & NV \\ 
& 31-08-2019  	&  2458727.46931	 & R  &  ~9,~45 & 0.73 & 2.83	& 1, ~8 & 0.01 & 11.26 & NV \\ 
& 22-10-2019  	&  2458779.40200	 & R  &  24,120 & 0.67 & 1.95	& 4, 20 & 1.36 & ~4.43 & NV \\ 
& 23-10-2019  	&  2458780.38514	 & R  &  17,~68 & 0.86 & 2.24	& 2, 12 & 6.88 & ~6.93 & NV \\ 
& 27-10-2019  	&  2458784.39414	 & R  &  24,120 & 1.34 & 1.95	& 4, 20 & 3.37 & ~4.43 & NV \\ 
& 04-11-2019  	&  2458792.51383	 & R  &  10,~50 & 1.53 & 2.70	& 1, ~8 & 0.06 & 11.26 & NV \\ 

1ES 0414$+$009
& 30-12-2016	& 2457753.10083 	& R &  34, 34 & 0.71 & 2.26 	&  ~6, 28 & 11.19 & 3.53 & NV  \\ 
& 19-01-2017	& 2457773.08978 	& R &  31, 31 & 0.21 & 2.35 	&  ~5, 24 & ~0.95 & 3.90 & NV  \\ 
& 31-10-2018	& 2458423.59324 	& R &  37, 37 & 1.94 & 2.18 	&  ~6, 28 & ~1.72 & 3.53 & NV  \\ 
& 02-11-2018	& 2458425.50561 	& R &  19, 19 & 2.03 & 3.03 	&  ~3, 16 & ~6.34 & 5.29 & NV  \\ 
& 15-12-2018	& 2458468.19977 	& R &  55, 55 & 0.20 & 1.89 	&  10, 44 & ~6.55 & 2.75 & NV  \\ 
& 16-12-2018	& 2458469.17221 	& R &  64, 64 & 0.30 & 1.80 	&  12, 52 & ~5.88 & 2.55 & NV  \\ 
& 28-12-2018	& 2458481.23748 	& R &  29, 29 & 0.48 & 2.42 	&  ~5, 24 & ~3.59 & 3.90 & NV  \\ 
& 29-12-2018	& 2458482.21406 	& R &  42, 42 & 0.23 & 2.08 	&  ~7, 32 & ~0.96 & 3.26 & NV  \\

1ES 2344$+$514
& 25-10-2016	& 2457687.09950 	& R &  58, 58 & 1.60 & 1.86 & 10, 44 & 19.55 & ~2.75 & NV \\ 
& 26-10-2016	& 2457688.08325 	& R &  69, 69 & 0.74 & 1.76 & 13, 56 & ~3.41 & ~2.47 & NV \\
& 08-11-2016	& 2457701.14346 	& R &  40, 40 & 0.15 & 2.11 & ~7, 32 & ~7.10 & ~3.26 & NV \\
& 09-11-2016	& 2457702.04397 	& R &  33, 33 & 0.50 & 2.29 & ~5, 24 & ~1.39 & ~3.90 & NV \\
& 24-11-2016	& 2457717.06291 	& R &  69, 69 & 1.53 & 1.76 & 13, 56 & ~8.49 & ~2.47 & NV \\
& 25-11-2016	& 2457718.04508 	& R &  59, 59 & 1.35 & 1.85 & 11, 48 & 13.47 & ~2.64 & NV \\ 
& 29-12-2016	& 2457752.09341 	& R &  35, 35 & 1.84 & 2.23 & ~6, 28 & ~8.02 & ~3.53 & NV \\
& 12-10-2017	& 2458039.14024 	& R &  ~9, ~9 & 0.69 & 5.35 & ~1, ~8 & ~2.70 & 11.26 & NV \\ 
& 10-10-2018	& 2458402.07662 	& R &  94, 94 & 2.12 & 1.62 & 18, 76 & ~8.86 & ~2.18 & ~V \\
& 15-10-2018	& 2458407.07281 	& R &  69, 69 & 1.33 & 1.76 & 13, 56 & ~1.82 & ~2.47 & NV \\
& 31-10-2018	& 2458423.33546 	& R &  25, 25 & 5.55 & 2.60 & ~4, 20 & ~3.38 & ~4.43 & NV \\ 
& 02-11-2018	& 2458425.30517 	& R &  63, 63 & 4.61 & 1.81 & 11, 48 & ~2.36 & ~2.64 & NV \\ 
& 15-12-2018	& 2458468.04470 	& R &  59, 59 & 1.68 & 1.85 & 11, 48 & ~7.68 & ~2.64 & NV \\
& 16-12-2018	& 2458469.02639 	& R &  59, 59 & 1.25 & 1.85 & 11, 48 & ~5.43 & ~2.64 & NV \\

\hline                          
\end{tabular}}
\end{table*}

\section{ANALYSIS TECHNIQUES} \label{sec:analysis}
To search for microvariations in the R-band differential light curves (DLCs) of the blazars 1ES 0229$+$200, 1ES 0414$+$009, and 1ES 2344$+$514, we employed two of the more recent and most reliable statistical tests: the power-enhanced {\it F}-test and the nested analysis of variance (ANOVA) test. Both the tests involve multiple comparison stars in the analysis, hence, are more powerful than the previously  frequently used $C-$test and $F-$test \citep{2014AJ....148...93D, 2015AJ....150...44D}. These statistical tests are described in details in our previous papers \citep[][and references therein]{2019ApJ...871..192P, 2020ApJ...890...72P}. Brief descriptions of these tests are given below.   

\begin{figure*}
\centering
\includegraphics[width=14cm, height=7cm]{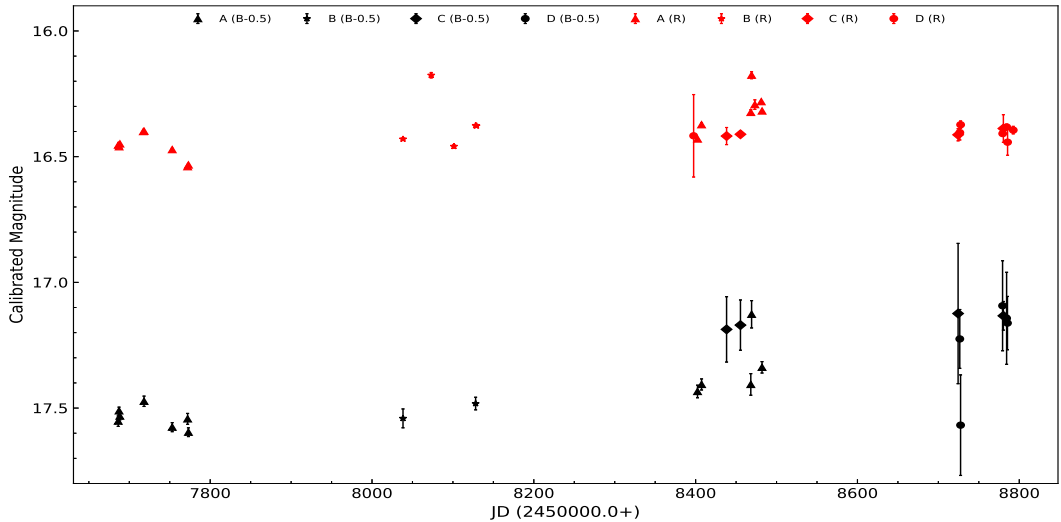} 
\caption{LTV optical ($BR$) light curves of 1ES  0229$+$200 shown in black ($B$) and red ($R$), respectively.  The telescopes employed are noted at the top of the figure.} 
\label{fig:ltv02}
\end{figure*}

\begin{figure}
\centering
\includegraphics[width=8.5cm, height=6cm]{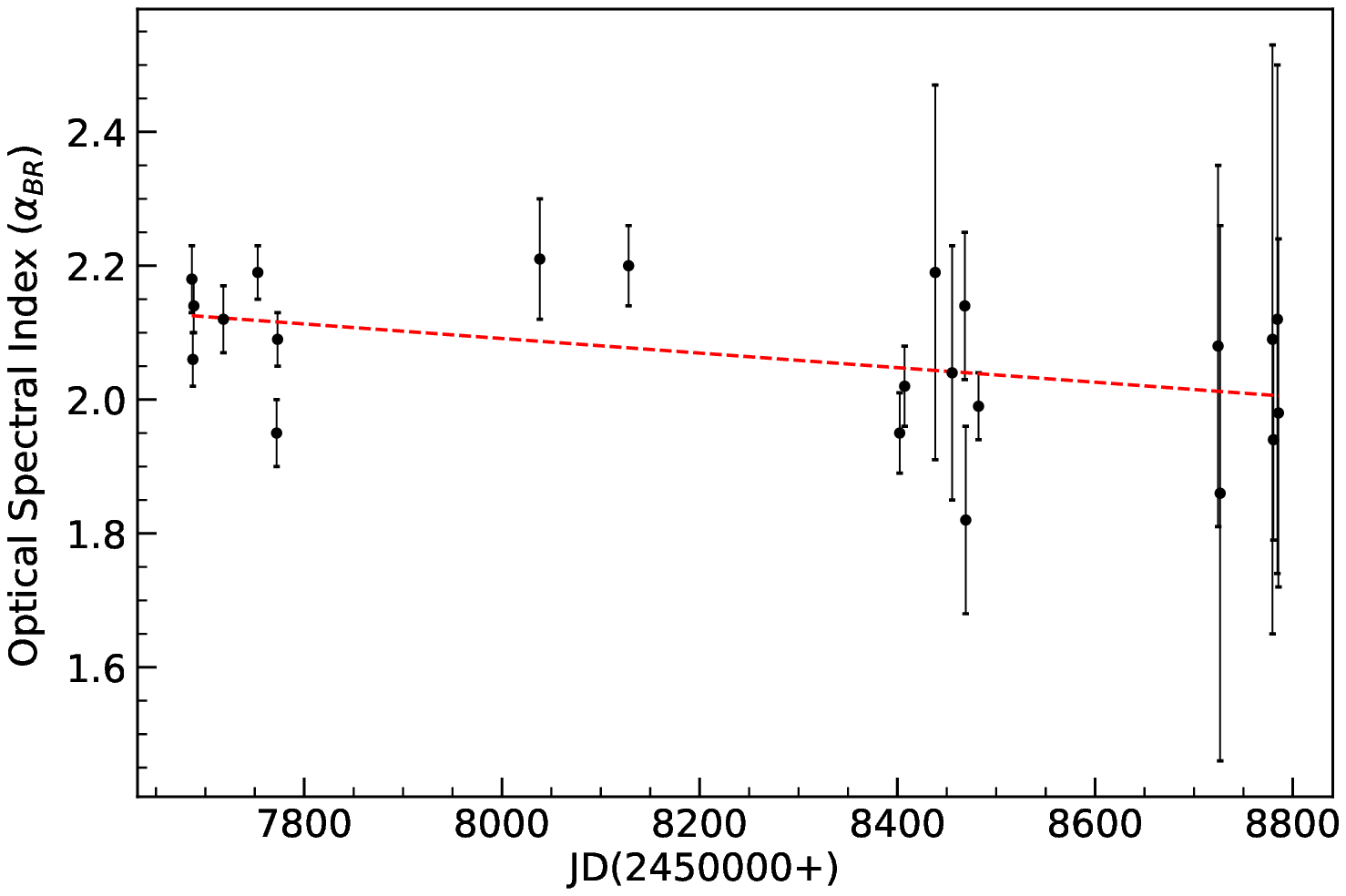}
\includegraphics[width=8.5cm, height=6cm]{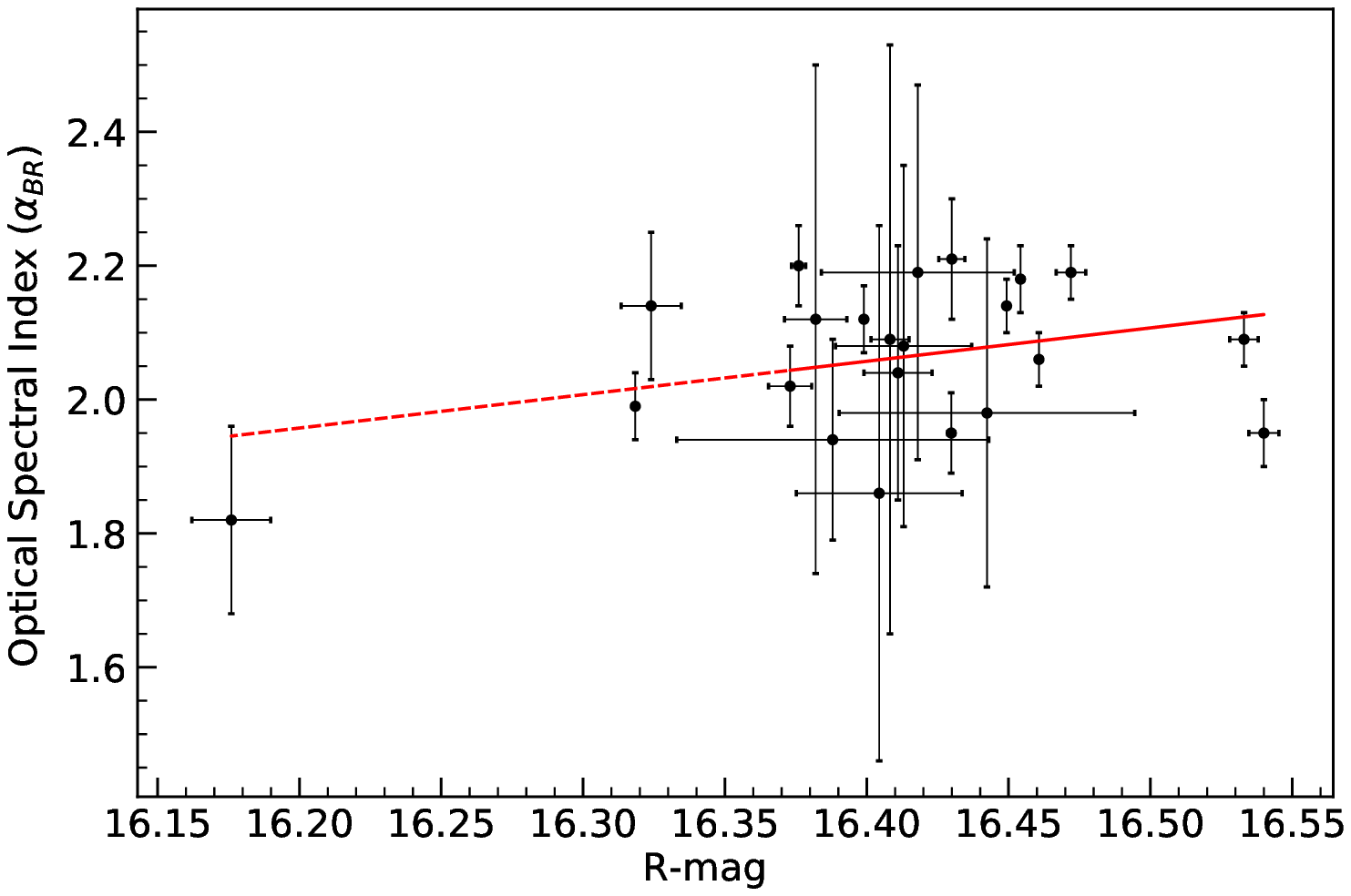}
\caption{Variation of optical spectral index ($\alpha_{BR}$) of 1ES 0229$+$200 with respect to time (top) and R-magnitude (bottom).} 
\label{fig:alpha02}
\end{figure}

\subsection{Power-enhanced {\it F}-test}\label{sec:f_test}
In the power-enhanced {\it F}-test, we use the brightest comparison star as a reference star to generate the DLCs of the blazar and the remaining ($k$) comparison stars. The power-enhanced {\it F}-test statistics is given by
\begin{equation} \label{eq:Fenh}
F_{\rm enh} = \frac{s_{\rm blz}^2}{s_c^2},
\end{equation}  
where $s_{\rm blz}^2$ is the variance of the blazar DLC  and $s_c^2$ is the combined variance of the DLCs of $k$ comparison stars. The value of $s_c^2$ is estimated as \citep{2019ApJ...871..192P}
\begin{equation}
s_c^2 = \frac{1}{(\sum_{j=1}^k N_j)-k}\sum_{j=1}^{k} \sum_{i=1}^{N_i} s_{j,i}^2,
\end{equation}
where $N_j$ is the number of data points of the $j$th comparison star and $s_{j,i}^2$ is the scaled square deviation for the $j$th comparison star defined as
\begin{equation}
s_{j,i}^2 = \omega_j(m_{j,i}-\bar{m_j})^2,
\end{equation}
where $\omega_j$, $m_{j,i}$, and $\bar{m_j}$ are the scaling factor, differential magnitude, and the mean magnitude of the $j$th comparison star DLC, respectively. The scaling factor $\omega_j$ is taken as the ratio of averaged square error of the blazar DLC to the averaged square error of the $j$th comparison star \citep{2011MNRAS.412.2717J}.

In the present work, we always observed two or more comparison stars in the blazar field. The blazar and all the comparison stars have the same number of observations ($N$). The degrees of freedom in the numerator, $\nu_1$, and denominator, $\nu_2$, in the power-enhanced {\it F} statistics are $N-1$, and $k(N-1)$, respectively. We estimated the F-statistics critical value, $F_c$, corresponding to  99 per cent confidence level and degrees of freedom $\nu_1$, and $\nu_2$. The value of $F_{\rm enh}$ is calculated from Equation \ref{eq:Fenh} and is compared with $F_c$. A light curve is called variable (V) if $F_{\rm enh} \geq F_c$; otherwise, we consider it to be nonvariable (NV).

\subsection{Nested {\it ANOVA}}\label{sec:anova}
The nested {\it ANOVA} test analysis involves all the comparison stars as reference stars to produce a set of DLCs of the blazar. Each of these DLCs is divided into several groups having five points in each group. We calculated the values of mean square due to groups, $MS_G$, and the mean square
due to nested observations in groups, $MS_{O(G)}$, following Equation (4) of \cite{2015AJ....150...44D}. The value of the {\it F}-statistic is then estimated as $F = MS_{G}/MS_{O(G)}$. A light curve is considered as variable if the value of $F-$statistic $\geq$ $F_c$ at 99 per cent confidence level, otherwise, we call it nonvariable.

\subsection{Flux Variability Amplitude}\label{sec:var_amp}
The flux variability amplitudes (Amp; in per cent) on IDV and LTV timescales for the extreme TeV blazars were calculated in the standard way as \citep{1996A&A...305...42H} 
\begin{equation}\label{eq:amp}
{\rm Amp} = 100\times \sqrt{(A_{\rm max}-A_{\rm min})^2 - 2 \sigma^2},
\end{equation}
where $A_{\rm max}$, $A_{\rm min}$, and $\sigma$ are the maximum magnitude, minimum magnitude, and the mean measurement error, respectively, in the calibrated light curve of the blazar. 

The results of IDV analyses of these extreme TeV blazars using both the statistical tests are given in Table \ref{tab:var_res1}. In the table, a light curve is conservatively declared as a variable only if statistically significant variations were found by both the probes, otherwise, we labeled it as nonvariable. The amplitude of variability is also mentioned in the last column of Table \ref {tab:var_res1} for the variable light curve.

\begin{table}
\caption{Variation of optical spectral index, $\alpha_{BR}$, with respect to time and R-magnitude of 1ES 0229$+$200} 
\label{tab:alpha_tr02} 
\centering 
\resizebox{0.5\textwidth} {!}{ 
\begin{tabular}{lcccc} 
\hline\hline 
Parameter & $m_1^a$ & $c_1^a$ & $r_1^a$ & $p_1^a$ \\
& & & & \\ 
\hline 
$\alpha_{BR}$ vs time  &$-1.09e-04 \pm 5.30e-05$ &   2.96 &  -0.42 & 5.33e-02 \\
$\alpha_{BR}$ vs R-mag &$  0.50 \pm   0.31$ &  -6.12 &   0.34 & 1.27e-01 \\

\hline 
\end{tabular}}\\
$^am_1$ = slope and $c_1$ = intercept of $\alpha_{BR}$ against time or R magnitude; $r_1$ = Correlation coefficient; $p_1$ = null hypothesis probability
\end{table}

\section{RESULTS}\label{sec:res}
\subsection{1ES 0229$+$200}
1ES 0229$+$200 ($\alpha_{\rm 2000} = 02^h32^m53.2^s;$ $\delta_{\rm 2000} = +20^{\circ}16^{\prime}21^{\prime\prime}$) was first detected in X-rays  with the \textit{Einstein} satellite's Imaging Proportional Counter (IPC) \citep{1992ApJS...80..257E} and later classified as a BL Lac object because of its featureless optical spectrum \citep{1993ApJ...412..541S}. It was listed as an HBL on the basis of its X-ray to radio flux ratio \citep{1995A&AS..109..267G}.  1ES 0229$+$200 is hosted by an elliptical galaxy at a  redshift of $z=0.1396$ \citep{2005ApJ...631..762W}. Optical ($BR$) observations of the blazar were performed between 2006 and 2010 with the ATOM telescope in a multiwavelength campaign by \cite{2011A&A...534A.130K}. Over five years of observing, they found no significant variations in the $R$ band; the average magnitudes in the $B$ and $R$ bands were $18.38 \pm 0.02$ and $16.39 \pm 0.01$, respectively. \cite{2015A&A...573A..69W} monitored the blazar 1ES 0229$+$200 for 184 nights during 2007--2012. They found  modest variations of $\sim$0.2 mag in its brightness and a very steep optical continuum with $\alpha_{BR}>3.7$.

We monitored the blazar 1ES 0229$+$200 between 2016 October 24 and 2019 April 4 for 31 nights during which a total of 534 image frames were taken in $B$ and $R$ wavebands. The complete photometric observation log of 1ES 0229$+$200 is given in Table \ref{tab:obs02}. 

\begin{figure*}
\centering
\includegraphics[width=14cm, height=7cm]{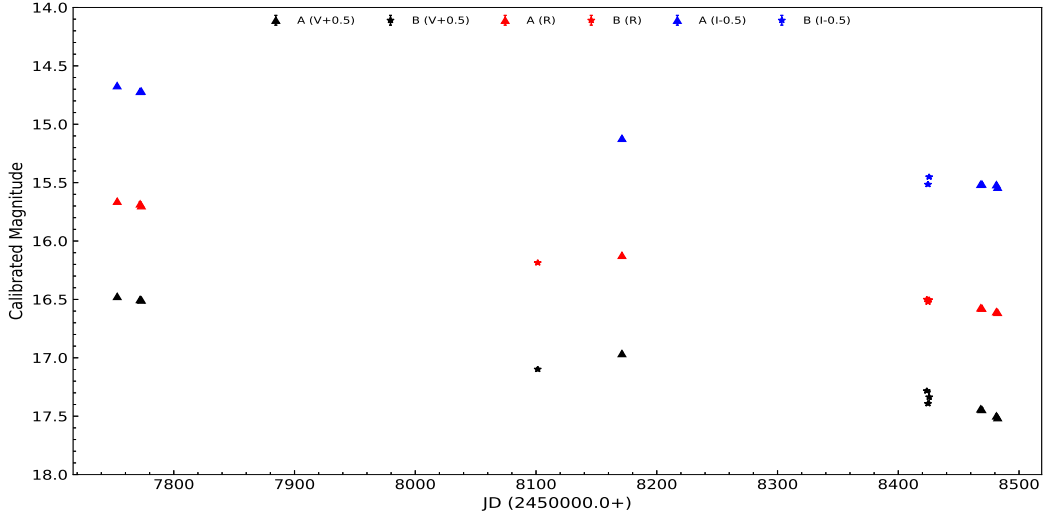} 
\caption{LTV optical ($VRI$) light curves of 1ES  0414$+$009; they are shown in black ($V$); red ($R$) and blue ($I$), respectively. The telescopes used are written at the top of the figure.} 
\label{fig:ltv04}
\end{figure*}

\subsubsection{Flux variability}
The calibrated $R-$band IDV light curves of the extreme TeV blazar 1ES 0229$+$200 are shown in Figure \ref{1_a} in Appendix \ref{app_c}. Visual inspection of these LCs indicates either no variation or very small fluctuations in the LCs over a few hours. We statistically examined these LCs for IDV using the tests discussed in Sections \ref{sec:f_test} and \ref{sec:anova}. The results of the IDV analysis are given in Table \ref{tab:var_res1}. No significant IDV was detected in any of these 14 LCs with enough data taken during a single night to conduct such a test.

The LTV-calibrated LCs of this source in optical $B$ and $R$ bands are shown in Figure \ref{fig:ltv02}, where we have plotted daily averaged magnitudes with respect to time. We shifted the $B$ band LC by -0.5 magnitude to make the LTV pattern more visible. The LTV LCs show variations in both the bands. The amplitudes of variability in $B$, and $R$ bands are 49.2 per cent, and 36.3 per cent, respectively. During our entire observing period, the blazar 1ES 0229$+$200 was in the brightest state of $R_{\rm mag} = 16.17$ on 2018 December 16, while the faintest magnitude detected was $R_{\rm mag} = 16.54$ on 2017 January 18. The average $B$ magnitude was 17.36 and the average $R$ magnitude was 16.39.

\subsubsection{Spectral variability}
To study the spectral variations of the TeV HBL 1ES 0229$+$200 on longer timescales, we calculated the $B-R$ color indices for the 22 nights with observations in both $B$ and $R$ bands. In the cases when there was more than one observation of the source during the same night, the average values were used. We then estimated the mean spectral indices, $\langle \alpha_{BR} \rangle$, of 1ES 0229$+$200 using the mean values of $B-R$ color indices, $\langle B - R \rangle$, as follows \citep{2015A&A...573A..69W},
\begin{equation}\label{eq:al_br}
\langle \alpha_{BR} \rangle = \frac{0.4 \langle B - R \rangle}{\mbox{log}(\nu_B/\nu_R)},
\end{equation}
where $\nu_B$ and $\nu_R$ are the effective frequencies of $B$ and $R$ bands, respectively. 

The value of spectral indices ($\alpha_{BR}$) ranges from $1.82 \pm 0.14$ to $2.21 \pm 0.09$. The weighted mean of the optical spectral index, $\alpha_{BR}$, was $2.09 \pm 0.01$. We plotted the spectral indices of 1ES 0229$+$200 with respect to time and $R-$band magnitude in the top and bottom panels of Figure \ref{fig:alpha02}, respectively. To search for any systematic variations in the spectral index, we fitted each panel in Figure \ref{fig:alpha02} with a polynomial of order 1. The results of the fits are given in Table \ref{tab:alpha_tr02}. No systematic temporal variation is found in the optical spectral index nor does the spectral index show any significant correlation with $R-$band magnitude.

\begin{figure}
\centering
\includegraphics[width=9cm, height=8cm]{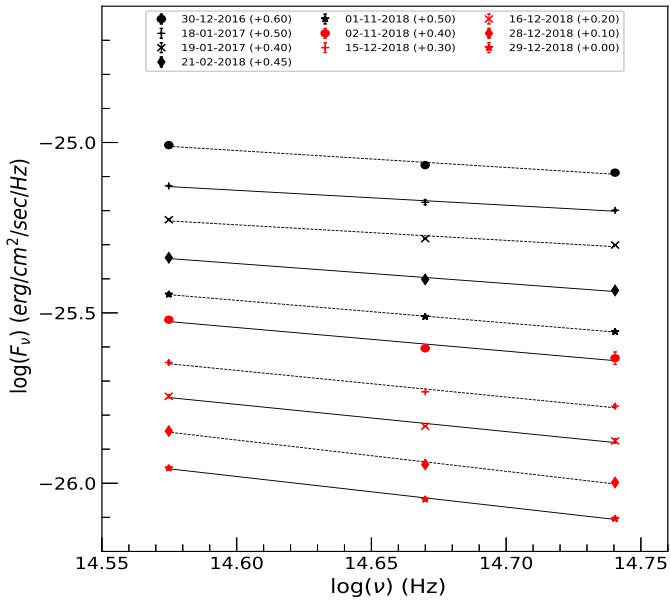}
\caption{\label{fig:sed04}Optical SEDs of 1ES 0414$+$009 in $V$, $R$, and $I$ bands.} 
\end{figure}

\begin{table*}
\caption{Results of a first order polynomial fits to optical SEDs of TeV blazars 1ES 0414$+$009 and 1ES 2344$+$514.}            
\label{tab:sed}                   
\centering     
\resizebox{0.7\textwidth} {!}{            
\begin{tabular}{lccccc}           
\hline\hline            		 
Blazar name  & Observation date & $\alpha^a$ &  $C^a$ & $r_2^a$ &  $p_2^a$  \\
	     &	dd-mm-yyyy      &            &          &         &                         \\		 
\hline        
1ES 0414$+$009	
& 30-12-2016	&$  0.50 \pm   0.08 $ & -17.79 &  -0.99 & 1.08e-01 \\
& 18-01-2017	&$  0.44 \pm   0.05 $ & -18.76 &  -0.99 & 7.03e-02 \\
& 19-01-2017	&$  0.46 \pm   0.09 $ & -18.57 &  -0.98 & 1.20e-01 \\
& 21-02-2018	&$  0.59 \pm   0.06 $ & -16.81 &  -1.00 & 6.29e-02 \\
& 01-11-2018	&$  0.66 \pm   0.02 $ & -15.77 &  -1.00 & 1.77e-02 \\
& 02-11-2018	&$  0.69 \pm   0.13 $ & -15.42 &  -0.98 & 1.20e-01 \\
& 15-12-2018	&$  0.78 \pm   0.09 $ & -14.27 &  -0.99 & 6.92e-02 \\
& 16-12-2018	&$  0.80 \pm   0.09 $ & -14.12 &  -0.99 & 6.93e-02 \\
& 28-12-2018	&$  0.92 \pm   0.08 $ & -12.47 &  -1.00 & 5.32e-02 \\
& 29-12-2018	&$  0.90 \pm   0.04 $ & -12.87 &  -1.00 & 3.09e-02 \\

1ES 2344$+$514	
& 06-09-2013	&$  1.35 \pm   0.09 $ &  -1.95 &  -1.00 & 4.40e-02 \\
& 20-10-2014	&$  1.45 \pm   0.18 $ &  -0.63 &  -0.99 & 7.69e-02 \\
& 06-11-2015	&$  1.41 \pm   0.22 $ &  -1.30 &  -0.99 & 9.90e-02 \\
& 12-11-2015	&$  1.28 \pm   0.23 $ &  -3.32 &  -0.98 & 1.13e-01 \\
& 16-11-2015	&$  1.13 \pm   0.34 $ &  -5.58 &  -0.96 & 1.85e-01 \\
& 24-10-2016	&$  1.51 \pm   0.07 $ &  -0.17 &  -1.00 & 3.02e-02 \\
& 25-10-2016	&$  1.49 \pm   0.11 $ &  -0.57 &  -1.00 & 4.62e-02 \\
& 26-10-2016	&$  1.50 \pm   0.08 $ &  -0.62 &  -1.00 & 3.38e-02 \\
& 08-11-2016	&$  1.56 \pm   0.13 $ &   0.17 &  -1.00 & 5.20e-02 \\
& 24-11-2016	&$  1.43 \pm   0.12 $ &  -1.78 &  -1.00 & 5.15e-02 \\
& 25-11-2016	&$  1.43 \pm   0.09 $ &  -1.90 &  -1.00 & 3.81e-02 \\
& 29-12-2016	&$  1.41 \pm   0.03 $ &  -2.28 &  -1.00 & 1.39e-02 \\
& 30-12-2016	&$  1.48 \pm   0.09 $ &  -1.34 &  -1.00 & 3.88e-02 \\
& 18-01-2017	&$  1.35 \pm   0.12 $ &  -3.31 &  -1.00 & 5.51e-02 \\
& 19-01-2017	&$  1.38 \pm   0.10 $ &  -3.07 &  -1.00 & 4.69e-02 \\
& 12-10-2017	&$  1.34 \pm   0.08 $ &  -3.74 &  -1.00 & 3.99e-02 \\
& 09-08-2018	&$  1.74 \pm   0.20 $ &   2.11 &  -0.99 & 7.28e-02 \\
& 10-09-2018	&$  1.31 \pm   0.26 $ &  -4.28 &  -0.98 & 1.23e-01 \\
& 11-09-2018	&$  1.31 \pm   0.12 $ &  -4.37 &  -1.00 & 6.00e-02 \\
& 12-09-2018	&$  1.07 \pm   0.63 $ &  -7.98 &  -0.86 & 3.39e-01 \\
& 05-10-2018	&$  1.36 \pm   0.17 $ &  -3.94 &  -0.99 & 7.76e-02 \\
& 11-10-2018	&$  1.43 \pm   0.10 $ &  -3.01 &  -1.00 & 4.39e-02 \\
& 15-10-2018	&$  1.41 \pm   0.10 $ &  -3.30 &  -1.00 & 4.38e-02 \\
& 31-10-2018	&$  1.52 \pm   0.05 $ &  -1.87 &  -1.00 & 2.23e-02 \\
& 01-11-2018	&$  1.44 \pm   0.10 $ &  -3.08 &  -1.00 & 4.63e-02 \\
& 02-11-2018	&$  1.49 \pm   0.08 $ &  -2.55 &  -1.00 & 3.50e-02 \\
& 02-12-2018	&$  1.32 \pm   0.01 $ &  -5.04 &  -1.00 & 6.46e-03 \\
& 15-12-2018	&$  1.44 \pm   0.12 $ &  -3.31 &  -1.00 & 5.38e-02 \\
& 16-12-2018	&$  1.41 \pm   0.11 $ &  -3.95 &  -1.00 & 5.01e-02 \\
& 28-12-2018	&$  1.45 \pm   0.14 $ &  -3.45 &  -1.00 & 6.02e-02 \\
& 29-12-2018	&$  1.37 \pm   0.14 $ &  -4.73 &  -0.99 & 6.58e-02 \\
& 06-06-2019	&$  1.39 \pm   0.04 $ &  -4.50 &  -1.00 & 2.04e-02 \\
& 07-07-2019	&$  1.59 \pm   0.07 $ &  -1.68 &  -1.00 & 2.98e-02 \\
& 31-08-2019	&$  1.55 \pm   0.08 $ &  -2.43 &  -1.00 & 3.31e-02 \\
& 23-10-2019	&$  1.58 \pm   0.03 $ &  -1.95 &  -1.00 & 1.14e-02 \\

\hline                           
\end{tabular}}\\
$^a\alpha$ = spectral index and $C$ = intercept of log($F_{\nu}$) against log($\nu$); $r_2$ = Correlation coefficient; $p_2$ = null hypothesis probability
\end{table*}

\begin{figure}
\centering
\includegraphics[width=8.5cm, height=6cm]{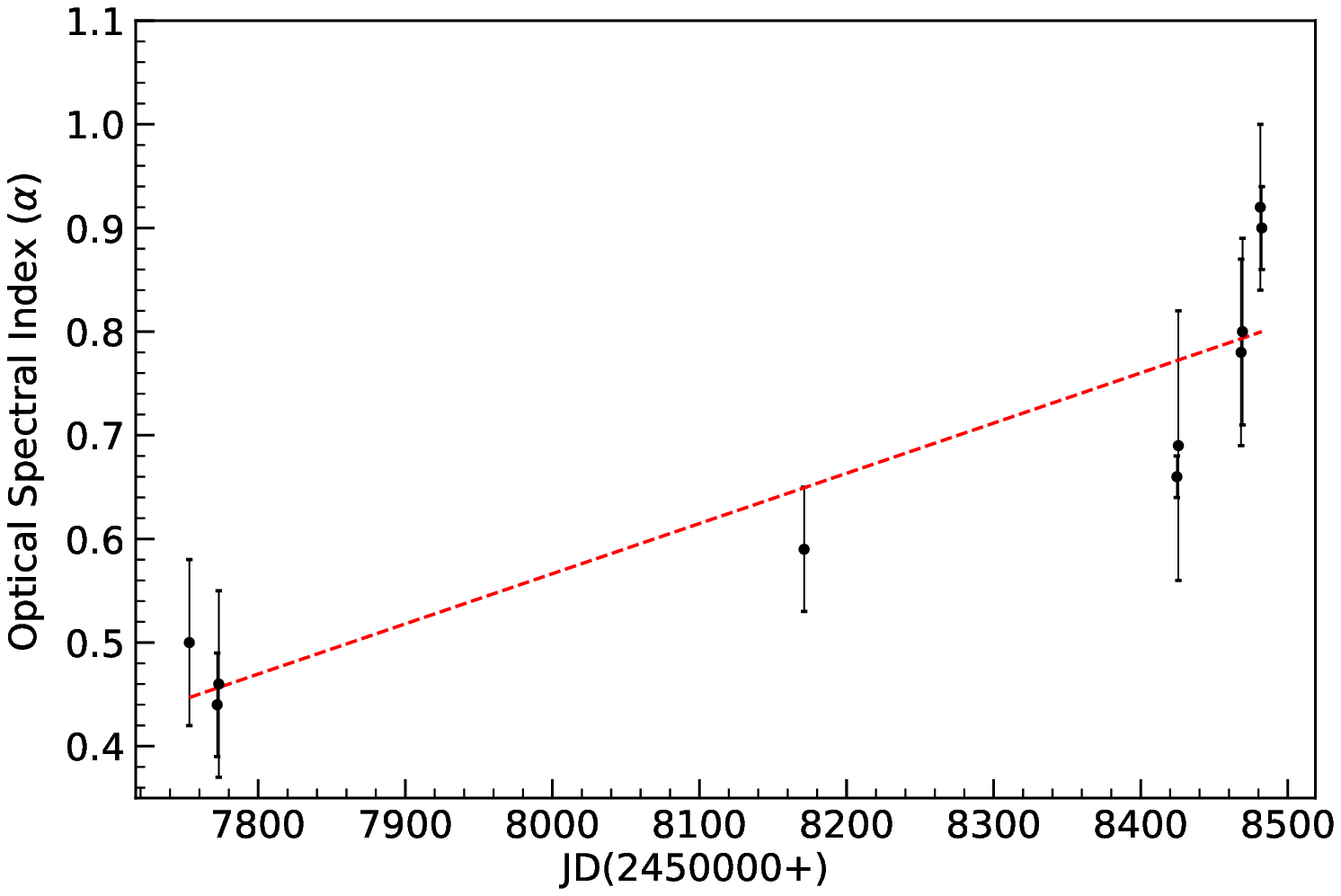}
\includegraphics[width=8.5cm, height=6cm]{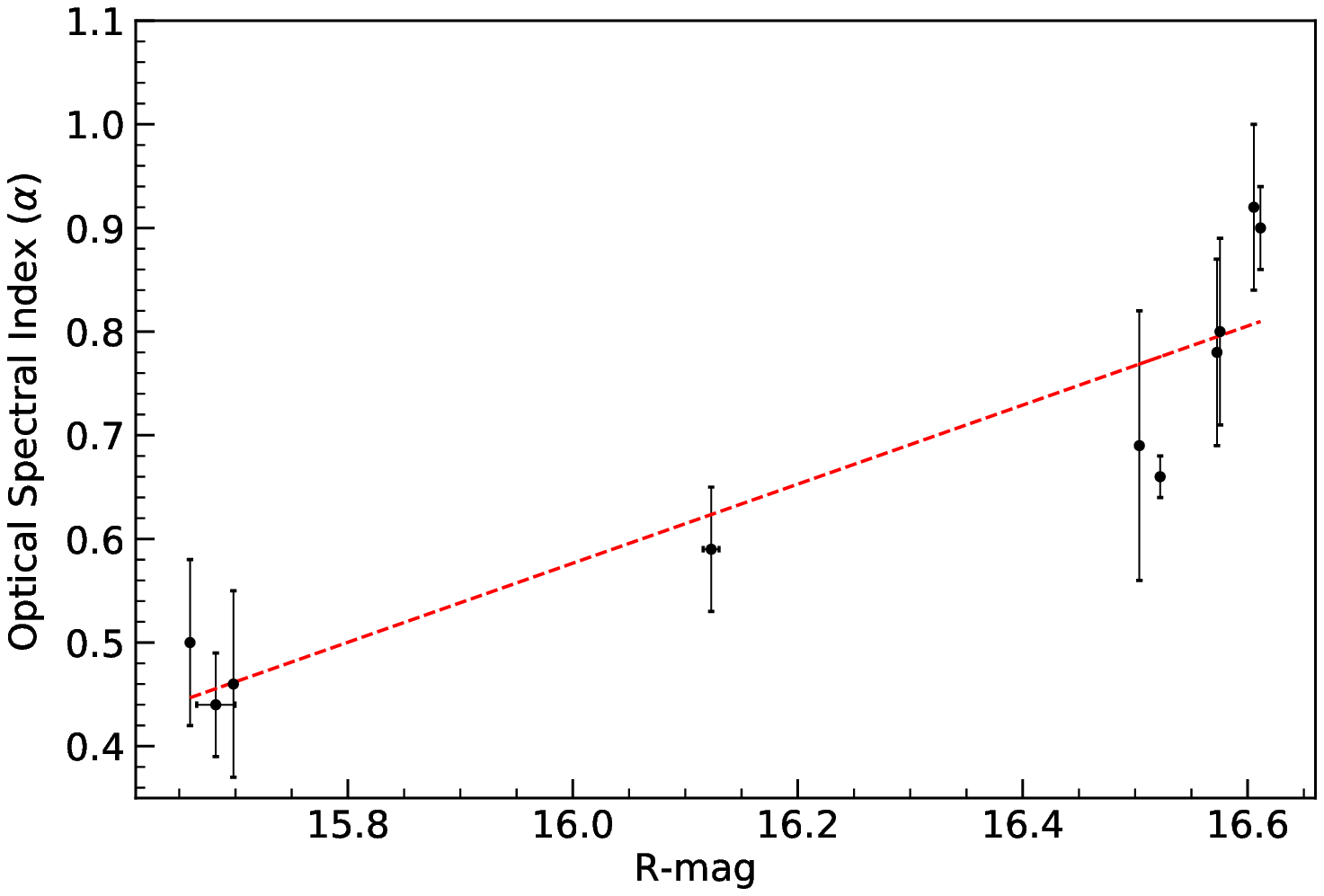}
\caption{Variation of optical spectral index of 1ES 0414$+$009 with respect to time (top) and R-magnitude (bottom).} 
\label{fig:alpha04}
\end{figure}

\begin{table}
\caption{Variation of optical spectral index, $\alpha$, with respect to time and R-magnitude during our observing campaign of 1ES 0414$+$009 and 1ES 2344$+$514.} 
\label{tab:alpha_tr} 
\centering 
\resizebox{0.5\textwidth} {!}{ 
\begin{tabular}{lccccc} 
\hline\hline 
Blazar & Parameter & $m_3^a$ & $c_3^a$ & $r_3^a$ & $p_3^a$ \\
	&& & & & \\ 
\hline 
1ES 0414$+$009  &$\alpha$ vs time  &$4.84e-04 \pm 8.10e-05$ &  -3.31 &   0.90 & 3.32e-04 \\
		&$\alpha$ vs R-mag &$  0.38 \pm   0.06$ &  -5.53 &   0.92 & 2.01e-04 \\

1ES 2344$+$514	&$\alpha$ vs time  &$4.71e-05 \pm 3.91e-05$ &   1.04 &   0.21 & 2.37e-01 \\
		&$\alpha$ vs R-mag &$ -0.51 \pm   0.26$ &   8.86 &  -0.33 & 5.66e-02 \\
\hline 
\end{tabular}}\\
$^am_3$ = slope and $c_3$ = intercept of $\alpha$ against time or R magnitude; $r_3$ = Correlation coefficient; $p_3$ = null hypothesis probability
\end{table}

\subsection{1ES 0414$+$009}
The blazar 1ES 0414$+$009 ($\alpha_{\rm 2000} = 04^h16^m53.0^s$; $\delta_{\rm 2000} = +01^{\circ}05^{\prime}20^{\prime\prime}$) was initially discovered as an X-ray source associated with a cluster of galaxies \citep{1980ApJ...235..351U} and later classified as a BL Lac object \citep{1983ApJ...270L...1U}. The redshift of 1ES 0414$+$009 is $z = 0.287$ \citep{1991AJ....101..818H}, derived from the weak stellar absorption lines. \cite{1992MNRAS.256..655M} reported $R = 16.64$ and $V = 17.21$ from observations during 1986 and 1987. 1ES 0414$+$009 was monitored from 1996 October -- 1997 February by \cite{1998A&AS..132..361R} with its $R$-mag ranging from 16.18 to 16.40.  \cite{2009MNRAS.398..832K} observed this BL Lac during 1997 October -- 2007 February on 56 nights. No IDV was found and the R brightness varied from 15.86 to 16.92 during this period. A similar brightness variation of $\leq$1 mag was reported from observations on 285 nights in 2007--2012 by \cite{2015A&A...573A..69W}, during which spectral indices of $\alpha_{BR} \sim 0.7-1.3$ were found.

We performed optical ($VRI$) photometric observations of this extreme blazar from 2016 December 30 to 2018 December 29 during 12 nights and observed a total of 346 image frames. The observation log of 1ES 0414$+$009 is given in Table \ref{tab:obs04}.

\begin{figure}
\centering
\includegraphics[width=8.5cm, height=6cm]{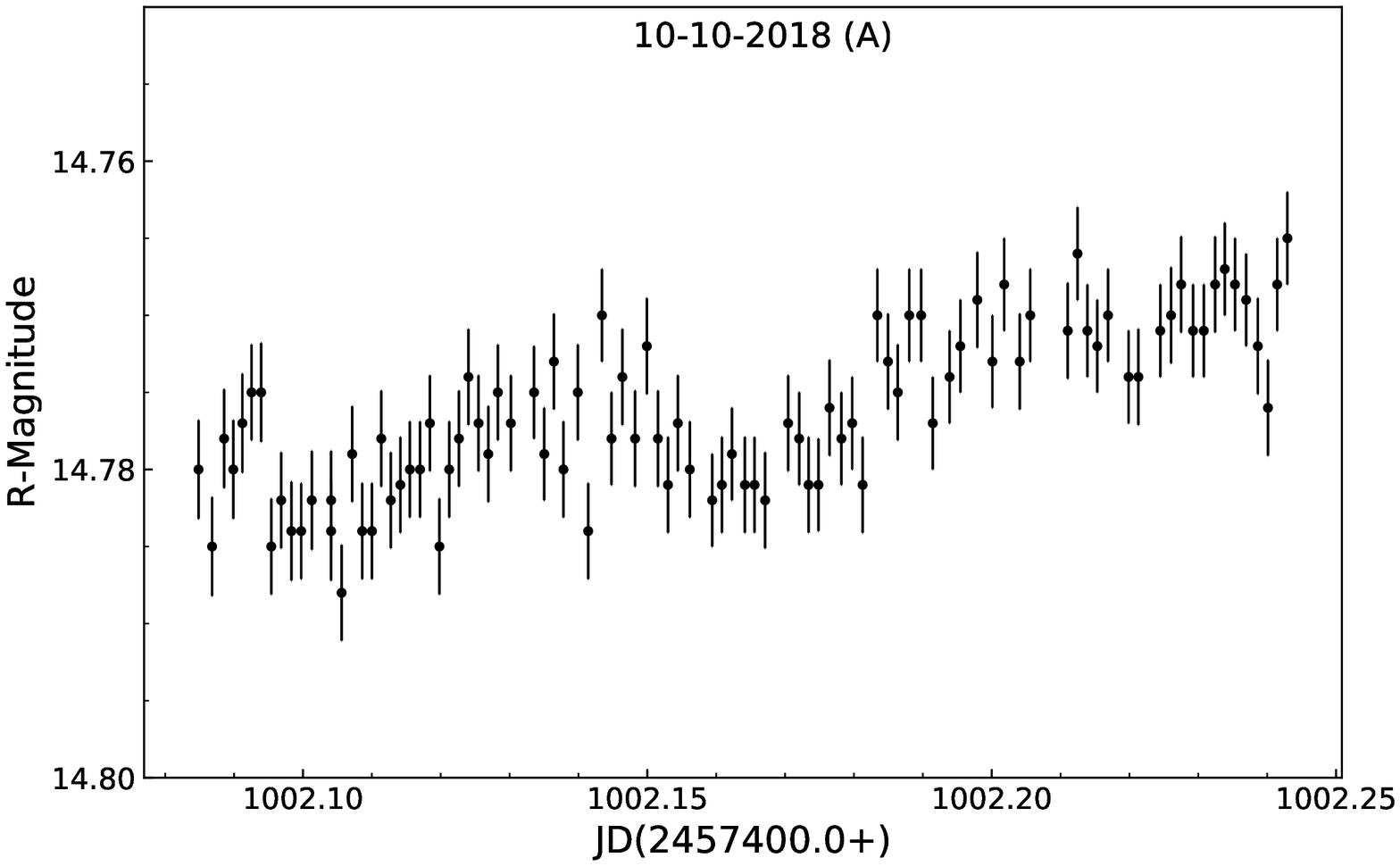}
\caption{\label{23a}The variable optical $R-$band IDV LC of the extreme TeV blazar 1ES 2344$+$514. The observation date and the telescope code are mentioned at the top.} 
\end{figure}
\begin{figure*}
\centering
\includegraphics[width=14cm, height=7cm]{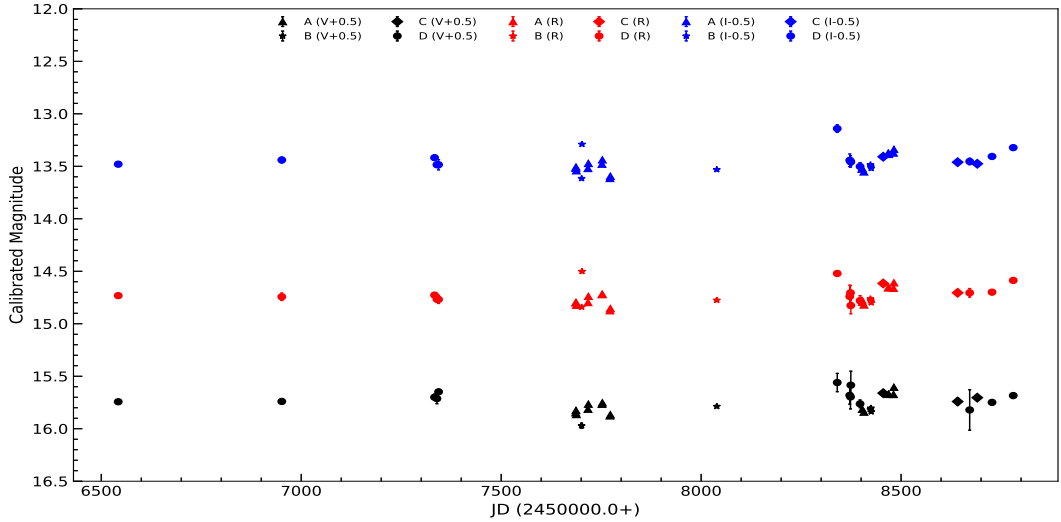} 
\caption{LTV optical ($VRI$) light curves of 1ES  2344$+$514; they are shown in black ($V$), red ($R$), and blue ($I$), respectively. The telescopes employed are written at the top of the figure.} 
\label{fig:ltv23}
\end{figure*}

\begin{figure*}
\centering
\includegraphics[width=14cm, height=16cm]{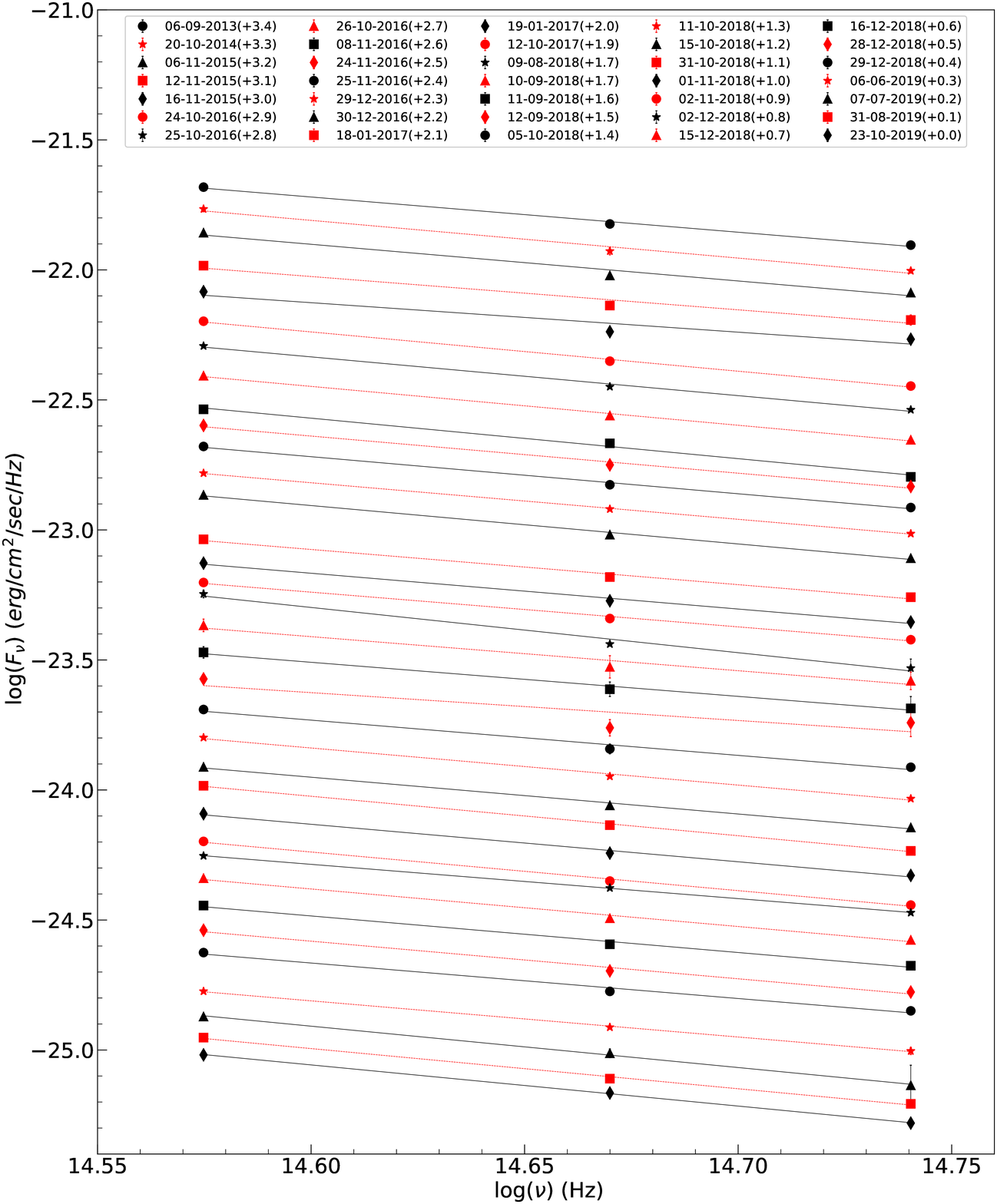}
\caption{Optical SEDs of 1ES 2344$+$514 in $V$, $R$, and $I$ bands.} 
\label{fig:sed23}
\end{figure*}

\begin{figure}
\centering
\includegraphics[width=8.5cm, height=6cm]{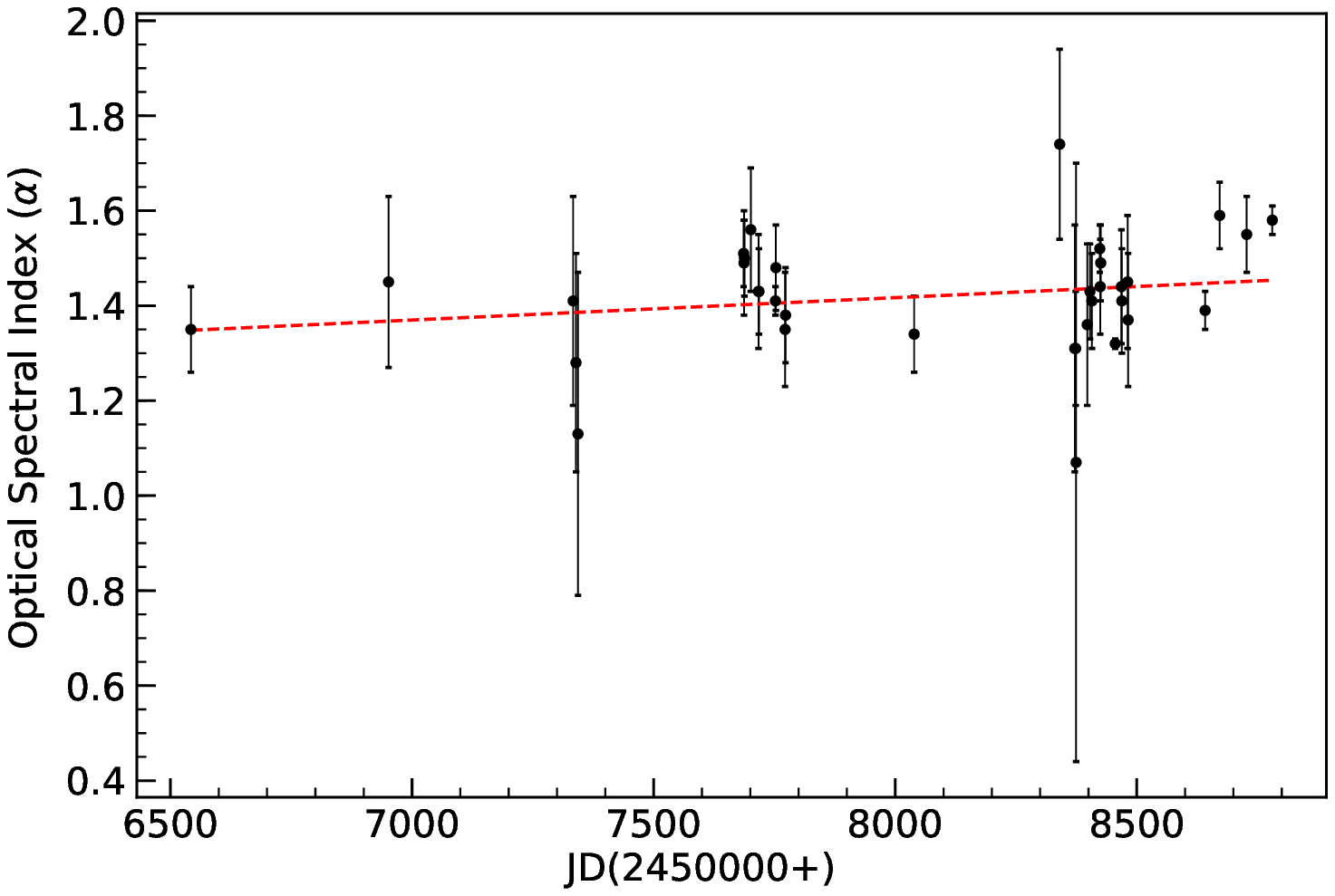}
\includegraphics[width=8.5cm, height=6cm]{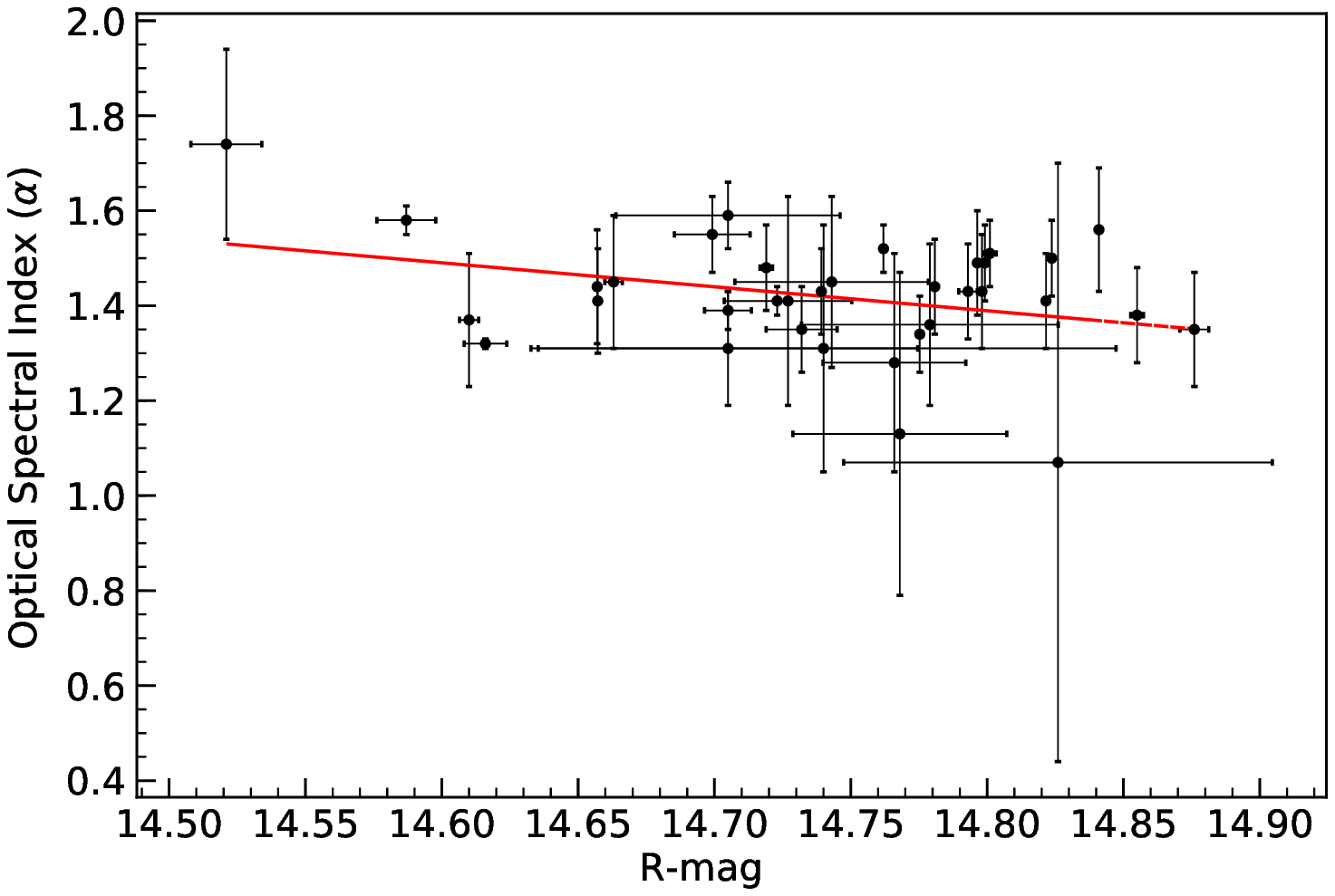}
\caption{Variation of optical spectral index of 1ES 2344$+$514 with respect to time (top) and R-magnitude (bottom).} 
\label{fig:alpha23}
\end{figure}

\subsubsection{Flux variability}
The calibrated IDV LCs of the TeV blazar 1ES 0414$+$009 in optical $R-$band are plotted in Figure \ref{1_b}. 
Using the statistical tests discussed in Sections \ref{sec:f_test} and \ref{sec:anova}, we found no evidence of statistically significant IDV in any of the 8 nights with 20 or more observations.

The LTV LCs of 1ES 0414$+$009 in $V$, $R$, and $I$ optical wavebands are shown in Figure \ref{fig:ltv04}, where nightly averaged calibrated magnitudes are plotted with respect to time. In the figure, the $V$, and $I$ band LCs are shifted by $+0.5$ mag, and $-0.5$ mag, respectively. Clear long-term variations are seen in all three optical bands. Using Equation \ref{eq:amp}, we have estimated the variability amplitudes of 103.8 per cent, 95.2 per cent, and 87.0 per cent, in $V$, $R$, and $I$ bands, respectively. During our observing campaign, the blazar 1ES 0414$+$009 was detected in the brightest state of $R_{\rm mag} = 15.66$ on 2016 December 30, while the faintest magnitude observed was $R_{\rm mag} = 16.61$ on 2018 December 29, a range very similar to those reported during earlier monitoring of this source \citep{2009MNRAS.398..832K,2015A&A...573A..69W}. The mean magnitudes over the two years of new data presented here were 16.62, 16.27, and 15.73 in the $V$, $R$, and $I$ bands, respectively.    

\subsubsection{Spectral variability}  
To study spectral variations during our observing period, we extracted the optical ({\it VRI}) SEDs of the blazar 1ES 0414$+$009 for 10 nights in which observations were performed in all three wavebands. For this, we first dereddened the calibrated $V$, $R$, and $I$ magnitudes by subtracting the Galactic extinction, $A_{\lambda}$, from them. The values of $A_{\lambda}$ were taken from the NASA Extragalactic Database (NED\footnote{\url{https://ned.ipac.caltech.edu/}}). The dereddened calibrated magnitudes in each band were then converted into corresponding flux densities, $F_{\nu}$. The optical SEDs of 1ES 0414$+$009, in log($\nu$) vs log($F_{\nu}$) representation, are plotted in Figure \ref{fig:sed04}.
 
Since a simple power law ($F_{\nu} \propto \nu^{-\alpha}$, where $\alpha$ is known as the optical spectral index) provides a better fit to the blazar optical continuum spectra, we fitted each SED of 1ES 0414$+$009 with a first order polynomial of the form log($F_{\nu}$) = $-\alpha$ log($\nu$)+ $C$ to get the values of optical spectral index. The results of the fits are given in Table \ref{tab:sed}.
The values of the spectral indices ($\alpha$) range from $0.44 \pm 0.05$ to $0.92 \pm 0.08$ and their weighted mean was $0.67 \pm 0.01$. We show the spectral indices of 1ES 0414$+$009 with respect to time and $R-$band magnitude in the top and bottom panels of Figure \ref{fig:alpha04}, respectively. We fitted each panel in Figure \ref{fig:alpha04} with a first order polynomial to investigate any systematic variations in the spectral index. The results of the fits are given in Table \ref{tab:alpha_tr}. The optical spectral index increases with time and it also shows significant positive correlations with $R-$band magnitude; however, it is clear that these are not independent correlations.

\subsection{1ES 2344$+$514}
1ES 2344$+$514 ($\alpha_{\rm 2000} = 23^h47^m04.8^s$; $\delta_{\rm 2000} = +51^{\circ}42^{\prime}17.9^{\prime\prime}$) was identified as a BL Lac object  at a redshift of $z=0.044$ \citep{1996ApJS..104..251P}. It was discovered as a TeV source with the Whipple Observatory $\gamma$-ray telescope, making it the third BL Lac object, after Mrk 421 and Mrk 501, to be detected at VHE \citep{1998ApJ...501..616C}. In the X-rays, rapid variability on a timescale of $\sim$ 5000s has been reported by \cite{2000MNRAS.317..743G}. 
\cite{1999bmtm.proc...20M} reported positive detection of microvariability in the optical LCs of 1ES 2344$+$514.  
\cite{2002MNRAS.329..689X} and \cite{2004ChJAA...4..133F}, separately, observed the blazar in 2000, but didn't find any significant IDV. A considerable IDV with the variability amplitude $\Delta R = 0.69 \pm 0.16$ on a timescale of $\Delta t = 4738$ was claimed to have been detected on 2005 December 28 by \cite{2010Ap&SS.327...35M}. Optical monitoring of the TeV BL Lac 1ES 2344$+$514 were carried out for 19 nights during 2009--2010 by \cite{2012MNRAS.420.3147G}. They found no significant variation on IDV timescales, but a brightness variation of $\Delta R \sim 0.65$ detected on LTV timescales. 

We observed the blazar 1ES 2344$+$514 for 38 nights between 2013 September 6 and 2019 October 23. During this period, we collected a total of 861 image frames in $V$, $R$, and $I$ bands. The optical photometric observation log of 1ES 2344$+$514 is given in Table \ref{tab:obs23}.

\subsubsection{Flux variability}
We statistically examined 14 optical $R-$band LCs of the extreme TeV blazar 1ES 2344$+$514 for intraday variations using the power-enhanced {\it F}-test and the nested ANOVA test. The results of the analysis are shown in Table \ref{tab:var_res1}. No statistically significant IDV was detected by both the tests during 13 of the 14 nights, the exception being 2018 October 10.  On that night, the amplitude of variability was a modest 2.26 per cent in the $R-$band LC. We plotted the only light curve showing variability in Figure \ref{23a} and all other non variable LCs in Figure \ref{1_c} in Appendix \ref{app_c}.

The LTV LCs of the blazar 1ES 2344$+$514 for our entire monitoring period are shown in Figure \ref{fig:ltv23}, where daily averaged $V$, $R$, and $I$ band calibrated magnitudes are plotted with respect to time. The amplitudes of variability in $V$, $R$, and $I$ bands are 40.9 per cent, 37.4  per cent, and 47.5 per cent, respectively. During our six year long observing program, the blazar 1ES 2344$+$514 was observed in the brightest state of $R = 14.50$ on 2016 November 9, while the faintest magnitude detected was $R = 14.88$ on 2017 January 18. The mean magnitudes were 15.25, 14.74, and 13.96 in $V$, $R$, and $I$ bands, respectively.

\subsubsection{Spectral variability}
Similar to our analysis of  1ES 0414$+$009, we extracted the 35 optical ($VRI$) SEDs of 1ES 2344$+$514 and present them in Figure \ref{fig:sed23}. We fitted these SEDs with the logarithm of a power law. The results of the fits are given in Table \ref{tab:sed}. The optical spectral index values ($\alpha$) range from $1.07 \pm 0.63$ to $1.74 \pm 0.20$. The weighted mean optical spectral index was $1.37 \pm 0.01$. The spectral indices are plotted against time in the top panel and $R-$band magnitude in the bottom panel of Figure \ref{fig:alpha23}. The results of linear fits to each panel in Figure \ref{fig:alpha23} are given in Table \ref{tab:alpha_tr}. The optical spectral index does not show significant systematic variation with  time, nor does it correlate with $R-$band magnitude.


\section{DISCUSSION AND CONCLUSION} \label{sec:diss}
In the present work we analyzed the optical photometric data for three extreme TeV blazars collected using four ground-based telescopes during 2013--2019. In particular, we studied the flux and spectral variability properties of these blazars in optical wavebands on both IDV and LTV timescales. We examined a total of 36 optical $R-$band IDV LCs of three blazars using two  powerful and robust statistical methods: the power-enhanced {\it F}-test and the nested ANOVA test. Only one of these 36 IDV LCs exhibits statistically significant intraday variations: 1ES 2344$+$514 on 2018 October 10, when the amplitude of variability was only 2.26 per cent. So from our IDV analysis, we conclude that optical LCs of these three TeV HBLs are either constant or show nominal variations on IDV timescales. 

Microvariations in blazar light curves are frequently detected and can be explained by the turbulent plasma flowing at relativistic speed in the jet \citep{2014ApJ...780...87M,2016ApJ...820...12P}. Different optical IDV behaviors have been observed in the LBLs and HBLs subclasses of blazars, with HBLs being relatively less variable in optical bands on IDV timescales \citep[e.g.][]{1998A&A...329..853H,2011MNRAS.416..101G}. Our results are in good agreement with this conclusion. 
The presence of a stronger magnetic field in HBLs could be responsible for the different optical microvariability behaviors of LBLs and HBLs \citep{1999A&AS..135..477R}. The stronger magnetic field in HBLs might prevent the development of small features (e.g. density inhomogeneities or bends) by Kelvin-Helmholtz instabilities in the bases of jets, which could otherwise interact with the shocks in jets to produce microvariability if its value is greater than the critical value $B_c$ given by \citep{1995Ap&SS.234...49R} 
\begin{equation}
B_c = \big[4\pi n_e m_e c^2(\Gamma^2 - 1)\big]^{1/2} \Gamma^{-1},
\end{equation}
where $n_e$ and $m_e$ are the density and rest mass of electron, respectively; $\Gamma$ is the jet flow's bulk Lorentz factor.

On longer timescales, all three blazars exhibited flux variations in all observed optical wavebands, as would be expected. Among these three blazars, 1ES 0414$+$009, showed the maximum variation, $\Delta R \sim 0.95$, during our monitoring period, which is in accord with the brightness variation range of $\leq 1$ mag as reported by \cite{2015A&A...573A..69W}. The brightness variation of $\Delta R \sim 0.37$ we detected in 1ES 0229$+$200 is slightly larger than the variation of $\Delta R \simeq 0.2$ observed between 2007 and 2012 by \cite{2015A&A...573A..69W}. The blazar 1ES 2344$+$514 showed a modest variation of $\Delta R \sim 0.38$ during our six years-long monitoring period. This variation is smaller than the variation of $\Delta R \sim 0.65$ detected by \cite{2012MNRAS.420.3147G} during 2009--2010. 

Flux variations on such longer timescales in the LCs of blazars can be reasonably explained by one or more of the following physical processes: acceleration of electrons to high energies by shock in relativistic jet plasma followed by subsequent cooling via different radiation processes \citep{1985ApJ...298..114M}, a change of the magnetic field \citep{2010ApJ...725.2344B}, and/or change in the Doppler factor caused by either helical motion of the emitting region within the jet or wiggling of jets or helical jets \citep[e.g.,][]{1992A&A...255...59C, 1992A&A...259..109G,1999A&A...347...30V}.

Using the $B - R$ color indices we estimated the spectral index, $\alpha_{BR}$, of 1ES 0229$+$200 and saw that it didn't show any correlation with time or the $R-$band mag. Comparing our results with those obtained by \cite{2015A&A...573A..69W}, we observed a relatively flat ($\alpha_{BR} \sim 1.8-2.2$)  optical continuum spectra of TeV blazar 1ES 0229$+$200. We also obtained the values of optical spectral indices for 1ES 0414$+$009, and 1ES 2344$+$514 by fitting a single power law in each of their optical ($VRI$) SEDs. The weighted mean values of $\alpha$ are 0.67, and 1.37 for 1ES 0414$+$009, and 1ES 2344$+$514, respectively. The optical spectral index, $\alpha$, of 1ES 0414$+$009 shows a significant positive correlation with $R-$mag indicating a bluer (or flatter)-when-brighter (BWB or FWB) trend; however, no such correlation is seen for 1ES 2344$+$514. The BWB trend is usually observed in BL Lacertae objects because their radiation is dominated by non-thermal jet emissions, while the redder (or steeper)-when-brighter (RWB or SWB) trend is more likely seen in FSRQs \citep[e.g.][]{2006A&A...450...39G,2015MNRAS.452.4263G}. However, no clear trend of this type has  been found by some authors \citep[e.g.,][]{2009ApJ...694..174B,2009ApJS..185..511P}. The BWB trend detected 1ES 0414$+$009 can be explained in different ways \citep{2004A&A...419...25F}. In the two-component model, the optical emission of a blazar is produced by one variable and another constant component where the variable component has a relatively flatter slope ($\alpha_{\rm var} < \alpha_{\rm const}$). Within a one component synchrotron model the BWB trend could be explained by the injection of fresh electrons resulting in an increased flux as the energy distribution of fresh electrons is harder in comparison to the cooled ones \citep{1998A&A...333..452K,2002PASA...19..138M}. 

Our optical photometric observations of three extreme TeV blazars reveal that their light curves are either constant or show small fluctuations on IDV timescales. However on longer timescales they exhibit variability in all optical bands. The extreme blazar 1ES 0414$+$009 showed the BWB trend, while the errors in the optical spectral indices of the other two blazars might have restricted us from measuring any trend, if present, in their spectral changes. To make any firm conclusions about the general nature of the optical variability of extreme TeV blazars a much larger sample of such blazars needs to be observed for even longer times, and preferably in more optical bands.     

\section*{ACKNOWLEDGEMENT}
We are grateful to the anonymous referee for his/her helpful comments/suggestions.
GD gratefully acknowledges the observing grant support from the Institute of Astronomy and NAO Rozhen, BAS, via bilateral joint research project ``Gaia Celestial Reference Frame (CRF) and fast variable astronomical objects" (the head is G.\ Damljanovic). This work is a part of the Projects no.\ 176011 ``Dynamics and kinematics of celestial bodies and systems", no.\ 176004 ``Stellar physics" and no.\ 176021 ``Visible and invisible matter in nearby galaxies: theory and observations" supported by the Ministry of Education, Science and Technological Development of the Republic of Serbia.
 
\bibliographystyle{mnras}
\bibliography{master}

\appendix
\section{TELESCOPES AND INSTRUMENTS DETAILS}\label{app_a}
\begin{table*}
\caption{Details of telescopes and instruments used}          
\label{tab:telescopes}                   
\centering   			 
\resizebox{1.\textwidth} {!}{                    
\begin{tabular}{lccccc}         		
\hline\hline 
Code                       &     A               &    \multicolumn{2}{c}   {B }             &         C           &      D              \\ \hline 
Telescope 	           & 1.3 m DFOT          & \multicolumn{2}{c} {1.04 m ST }          & 1.4 m ASV           & 60 cm ASV            \\ \cmidrule[0.02cm](r){3-4}
CCD Model   		   & Andor 2K CCD	 & Tektronics 1K CCD  &  PyLoN CCD          & Andor iKon-L        & Apogee Alta  E47	  \\
Chip Size (pixels) 	   & 2048 $\times$ 2048  & 1024 $\times$ 1024 &  1340 $\times$ 1300 & 2048 $\times$ 2048  & $ 1024 \times 1024 $   \\
Pixel Size ($\mu m$)       & 13.5 $\times$ 13.5  & $24 \times 24$     &  $20 \times 20$     & 13.5 $\times$ 13.5  & $ 13 \times 13 $	  \\
Scale (arcsec/pixel)   	   & 0.535		 & 0.37               &  0.305	 	    & 0.244               & 0.447 	          \\
Field (arcmin$^2$)    	   & $18 \times 18$	 & $ 6 \times 6 $     &  $6.8 \times 6.6$   & 8.3 $\times$ 8.3    & $ 7.6 \times 7.6 $	  \\ 
Gain ($e^-$/ADU)	   & 2.0		 & 11.98              &  4.0	 	    & 1.0                 & 3.5	                   \\ 
Read-out Noise ($e^-$ rms) & 7.0		 & 6.9                &  6.4	 	    & 7.0                 & 10	                   \\	 
Typical Seeing (arcsec)    & 1.2--2.0		 & 1.4--2.6           &  1.2--2.1	    & 1.0-1.5             & 1.0-1.5	          \\ \hline                         
\end{tabular}}
\justify
Note. A: 1.3 m Devasthal Fast Optical Telescope (DFOT) at ARIES, Nainital, India. 
B: 1.04 m Sampuranand Telescope (ST) at ARIES, Nainital, India.
C: 1.4 m Milankovi\'{c} telescope at Astronomical Station Vidojevica (ASV), Serbia. 
D: 60 cm Nedeljkovi\'{c} telescope at Astronomical Station Vidojevica (ASV), Serbia. \\
\end{table*}

\section{OBSERVATION LOGS}\label{app_b}

\begin{table}
\caption{Observation log of optical photometric observations of 1ES 0229$+$200. See Table \ref{tab:telescopes} for telescope codes.}            
\label{tab:obs02}                   
\centering     
\resizebox{0.5\textwidth} {!}{                  
\begin{tabular}{ccccccc}           
\hline\hline               	
Obs. date       & Obs. start time  & Telescope  & Data points \\
dd$-$mm$-$yyyy  &       JD         &            & B,V,~R,I   \\\hline
24$-$10$-$2016  &  2457686.30358   &    A       & 1,0,40,0 \\
25$-$10$-$2016  &  2457687.31715   &    A       & 1,0,32,0 \\
26$-$10$-$2016  &  2457688.30863   &    A       & 1,0,30,0 \\
24$-$11$-$2016  &  2457717.28042   &    A       & 0,0,41,0 \\
25$-$11$-$2016  &  2457718.22020   &    A       & 1,0,50,0 \\
30$-$12$-$2016  &  2457753.05468   &    A       & 1,0,~1,0 \\
18$-$01$-$2017  &  2457772.08349   &    A       & 1,0,~1,0 \\
19$-$01$-$2017  &  2457773.06619   &    A       & 1,0,~1,0 \\
11$-$10$-$2017  &  2458038.30098   &    B       & 1,0,~1,0 \\
15$-$11$-$2017  &  2458073.15895   &    B       & 0,0,~1,0 \\
13$-$12$-$2017  &  2458101.24158   &    B       & 0,0,~1,0 \\
09$-$01$-$2018  &  2458128.06562   &    B       & 1,0,~1,0 \\
10$-$01$-$2018  &  2458129.06628   &    B       & 0,0,~1,0 \\
06$-$10$-$2018  &  2458397.52111   &    D       & 0,0,~1,0 \\
10$-$10$-$2018  &  2458402.32068   &    A       & 1,0,42,0 \\
15$-$10$-$2018  &  2458407.33545   &    A       & 1,0,~1,0 \\ 
15$-$11$-$2018  &  2458438.24477   &    C       & 1,0,~1,0 \\ 
02$-$12$-$2018  &  2458455.39085   &    C       & 1,0,~1,0 \\
15$-$12$-$2018  &  2458468.18387   &    A       & 1,0,~1,0 \\
16$-$12$-$2018  &  2458469.16480   &    A       & 1,0,~1,0 \\
20$-$12$-$2018  &  2458473.17715   &    A       & 0,0,~1,0 \\
28$-$12$-$2018  &  2458481.05990   &    A       & 0,0,34,0 \\
29$-$12$-$2018  &  2458482.05153   &    A       & 1,0,47,0 \\
28$-$08$-$2019  &  2458724.42101   &    C       & 1,0,48,0 \\
30$-$08$-$2019  &  2458726.50499   &    D       & 1,0,~7,0 \\ 
31$-$08$-$2019  &  2458727.46931   &    D       & 0,0,10,0 \\
22$-$10$-$2019  &  2458779.40200   &    D       & 1,0,25,0 \\ 
23$-$10$-$2019  &  2458780.38514   &    C       & 1,0,18,0 \\ 
27$-$10$-$2019  &  2458784.39414   &    D       & 1,0,25,0 \\ 
28$-$10$-$2019  &  2458785.43900   &    D       & 1,0,~7,0 \\ 
04$-$11$-$2019  &  2458792.51383   &    D       & 0,0,11,0 \\    
\hline              
\end{tabular}}\\
\end{table}

\begin{table}
\caption{Observation log of optical photometric observations of 1ES 0414$+$009. See Table \ref{tab:telescopes} for telescope codes.}          
\label{tab:obs04}                   
\centering     
\resizebox{0.5\textwidth} {!}{                  
\begin{tabular}{cccccc}           
\hline\hline               	
  Obs. date     &  Obs. start time & Telescope  & Data points \\
dd$-$mm$-$yyyy  &       JD         &            &  B,V,~R,I   \\\hline
30$-$12$-$2016  &  2457753.09527   &    A       & 0,1,35,1 \\
18$-$01$-$2017  &  2457772.11041   &    A       & 0,2,~2,1 \\ 
19$-$01$-$2017  &  2457773.09260   &    A       & 0,1,32,1 \\
13$-$12$-$2017  &  2458101.40040   &    B       & 0,1,~1,0 \\
21$-$02$-$2018  &  2458171.16218   &    A       & 0,1,~1,1 \\  
31$-$10$-$2018  &  2458423.62996   &    B       & 0,1,38,0  \\	
01$-$11$-$2018  &  2458424.58669   &    B       & 0,1,~1,1 \\ 
02$-$11$-$2018  &  2458425.59105   &    B       & 0,1,20,1 \\
15$-$12$-$2018  &  2458468.23930   &    A       & 0,1,56,1 \\ 
16$-$12$-$2018  &  2458469.18488   &    A       & 0,1,65,1 \\
28$-$12$-$2018  &  2458481.30444   &    A       & 0,1,30,1  \\
29$-$12$-$2018  &  2458482.29254   &    A       & 0,1,43,1  \\  
\hline              
\end{tabular}}\\
\end{table}

\begin{table}
\caption{Observation log of optical photometric observations of 1ES 2344$+$514. See Table \ref{tab:telescopes} for telescope codes.}           
\label{tab:obs23}                   
\centering     
\resizebox{0.5\textwidth} {!}{                  
\begin{tabular}{cccccc}           
\hline\hline               	
Obs. date       &  Obs. start time & Telescope  & Data points \\
dd$-$mm$-$yyyy  &       JD         &            & B,V,~R,I   \\\hline
06$-$09$-$2013  &  2456542.41798   &    D       & 0,1,~1,1  \\
20$-$10$-$2014  &  2456951.43801   &    D       & 0,1,~1,1 \\
06$-$11$-$2015  &  2457333.43866   &    D       & 0,1,~1,1 \\
12$-$11$-$2015  &  2457339.35205   &    D       & 0,1,~1,1 \\
16$-$11$-$2015  &  2457343.47328   &    D       & 0,1,~1,1 \\
24$-$10$-$2016  &  2457686.27076   &    A       & 0,1,~1,1 \\	 
25$-$10$-$2016 	&  2457687.08169   &    A       & 0,1,59,1 \\
26$-$10$-$2016  &  2457688.07811   &    A       & 0,1,70,1 \\
08$-$11$-$2016 	&  2457701.13915   &    B       & 0,1,41,1 \\
09$-$11$-$2016  &  2457702.04397   &    B       & 0,0,34,1 \\
24$-$11$-$2016  &  2457717.06578   &    A       & 0,1,70,1 \\
25$-$11$-$2016  &  2457718.04218   &    A       & 0,1,60,1 \\
29$-$12$-$2016  &  2457752.08833   &    A       & 0,1,36,1 \\
30$-$12$-$2016  &  2457753.04718   &    A       & 0,1,~1,1 \\
18$-$01$-$2017  &  2457772.09645   &    A       & 0,1,~1,1 \\
19$-$01$-$2017  &  2457773.05658   &    A       & 0,1,~1,1 \\  
12$-$10$-$2017  &  2458039.13509   &    B       & 0,1,10,1 \\ 
09$-$08$-$2018  &  2458340.35756   &    D       & 0,1,~1,1 \\
10$-$09$-$2018  &  2458371.60309   &    D       & 0,1,~1,1 \\
11$-$09$-$2018  &  2458373.45343   &    D       & 0,1,~1,1 \\
12$-$09$-$2018  &  2458374.35734   &    D       & 0,1,~1,1 \\
05$-$10$-$2018  &  2458397.43260   &    D       & 0,1,~1,1 \\ 
10$-$10$-$2018  &  2458402.07662   &    A       & 0,0,95,1 \\ 
11$-$10$-$2018  &  2458403.11768   &    A       & 0,1,~1,1 \\
15$-$10$-$2018  &  2458407.12018   &    A       & 0,1,70,1 \\ 
31$-$10$-$2018  &  2458423.43715   &    B       & 0,1,26,1 \\
01$-$11$-$2018  &  2458424.34072   &    B       & 0,1,~1,1 \\
02$-$11$-$2018  &  2458425.34785   &    B       & 0,1,64,1 \\
02$-$12$-$2018  &  2458455.35558   &    C       & 0,1,~1,1 \\	
15$-$12$-$2018  &  2458468.06682   &    A       & 0,1,60,1 \\
16$-$12$-$2018  &  2458469.04826   &    A       & 0,1,60,1 \\
28$-$12$-$2018  &  2458481.04459   &    A       & 0,1,~1,1  \\
29$-$12$-$2018  &  2458482.03729   &    A       & 0,1,~1,1  \\ 
06$-$06$-$2019  &  2458641.46473   &    C       & 0,1,~1,1 \\
07$-$07$-$2019  &  2458671.52386   &    D       & 0,1,~1,1  \\
26$-$07$-$2019  &  2458690.52772   &    C       & 0,1,~0,1 \\
31$-$08$-$2019  &  2458727.39598   &    D       & 0,6,~6,6 \\
23$-$10$-$2019  &  2458780.44473   &    D       & 0,6,~6,6 \\ 
\hline              
\end{tabular}}\\
\end{table}

\section{R-BAND INTRADAY LIGHT CURVES}\label{app_c}

\begin{figure*}
\centering
\includegraphics[width=18cm, height=16cm]{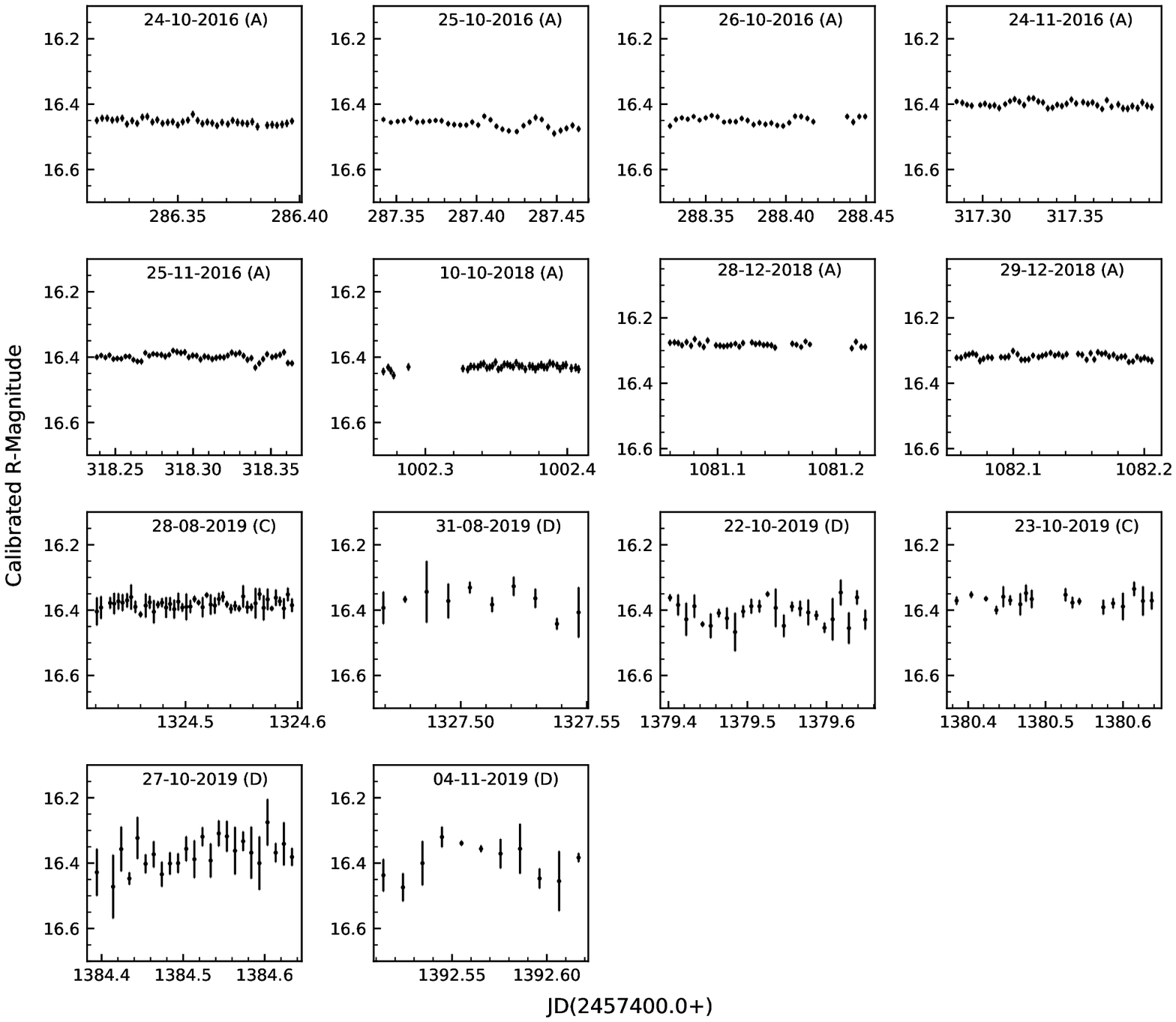}
\caption{\label{1_a}Optical $R-$band IDV LCs of the extreme TeV blazar 1ES 0229$+$200. The observation date and the telescope code are given in each plot.} 
\end{figure*}

\begin{figure*}
\centering
\includegraphics[width=18cm, height=8cm]{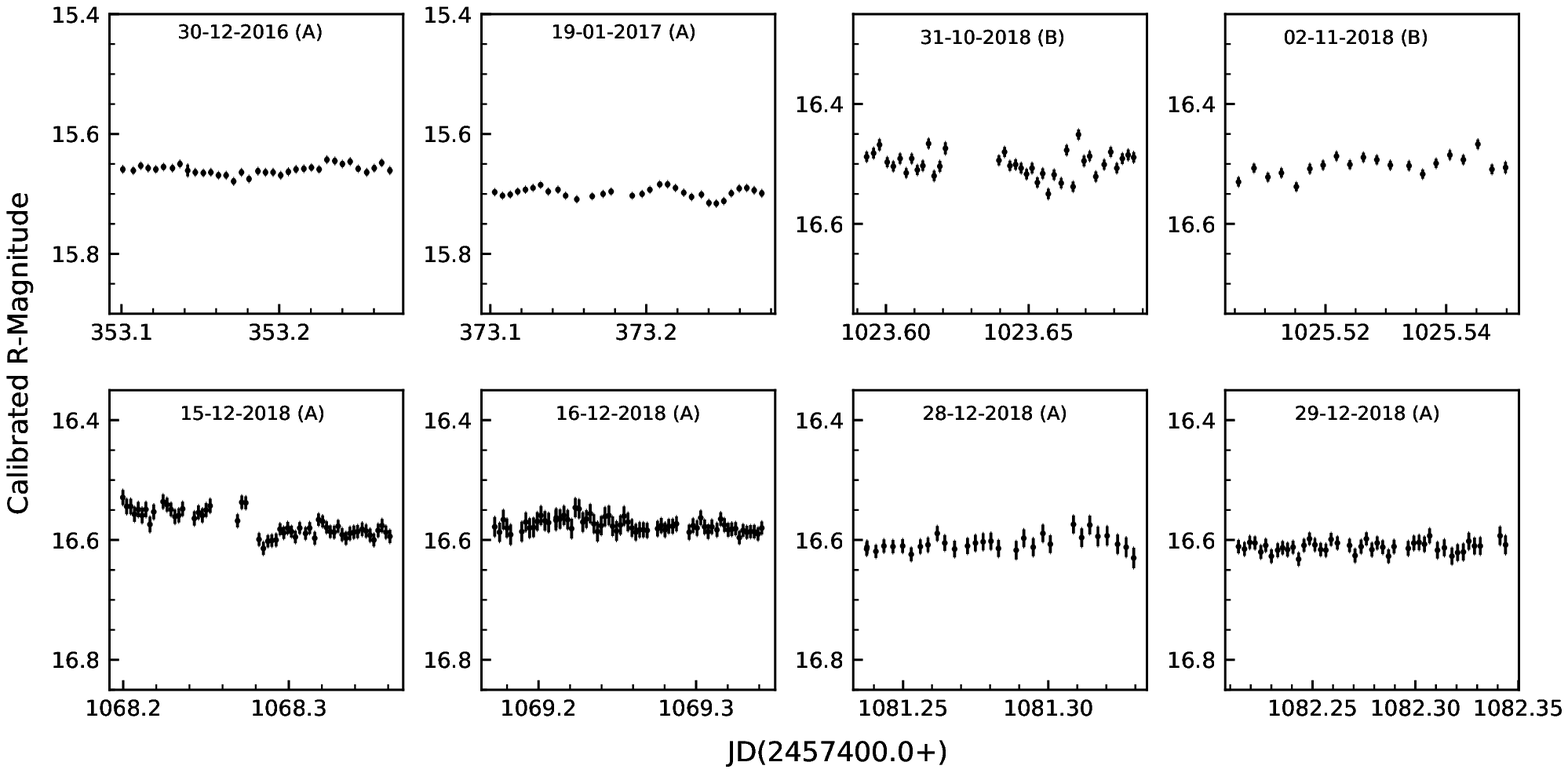}
\caption{\label{1_b}Optical $R-$band IDV LCs of the extreme TeV blazar 1ES 0414$+$009. The observation date and the telescope code are given in each plot.} 
\end{figure*}

\begin{figure*}
\centering
\includegraphics[width=18cm, height=16cm]{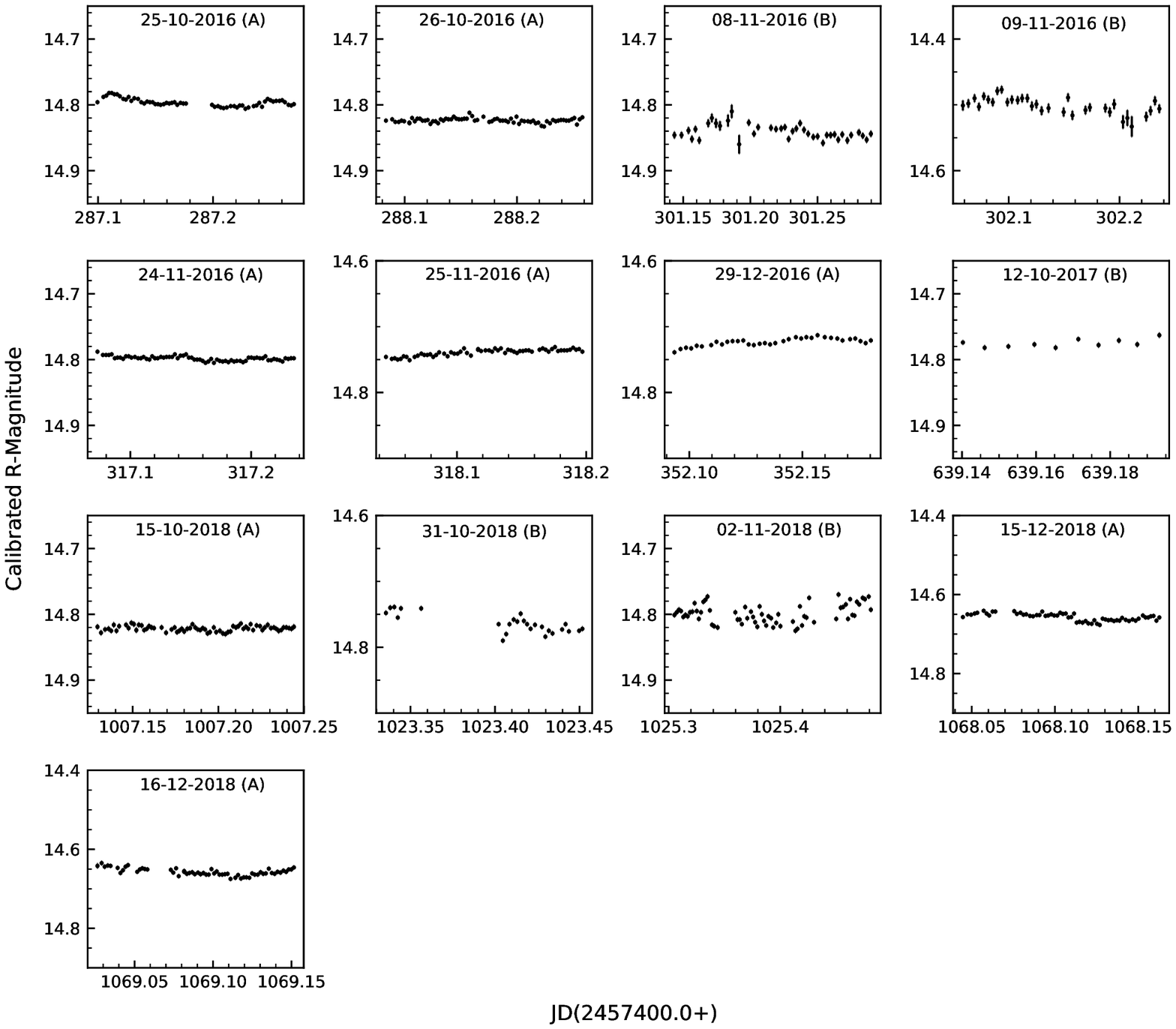}
\caption{\label{1_c}Optical $R-$band IDV LCs of the extreme TeV blazar 1ES 2344$+$514. The observation date and the telescope code are given in each plot.} 
\end{figure*}

\bsp	
\label{lastpage}
\end{document}